\documentclass[twocolumn]{aastex701}

\usepackage{hyperref}
\usepackage{enumitem}
\usepackage{amssymb}
\usepackage{lipsum}
\usepackage{subcaption}
\usepackage{placeins}
\usepackage{url}

\providecommand{\sorthelp}[1]{}

\newcommand{\um}{\mu\mathrm{m}}
\newcommand{\spherex}{SPHEREx}
\newcommand{\planck}{Planck}
\newcommand{\skysim}{Simulator}
\newcommand{\healpix}{HEALPix}
\newcommand{\w}{\mathrm{W}/\mathrm{m}^{2}/\mathrm{sr}}
\newcommand{\MJysr}{\mathrm{MJy}/\mathrm{sr}}

\newcommand{\HII}{H\,{\sc ii}}
\newcommand{\bralpha}{Br$\alpha$}
\newcommand{\paalpha}{Pa$\alpha$}
\newcommand{\pfbeta}{Pf$\beta$}
\newcommand{\wlmin}{0.75}
\newcommand{\wlmax}{5.0}
\newcommand{\nside}{N_\mathrm{side}}
\newcommand{\figref}[1]{Figure~\ref{#1}}
\newcommand{\secref}[1]{Section~\ref{#1}}

\newdimen\saa  \newdimen\sbb
\def\arcsec{\ifmmode {^{\scriptstyle\prime\prime}}
          \else $^{\scriptstyle\prime\prime}$\fi}
\def\arcmin{\ifmmode {^{\scriptstyle\prime}}
          \else $^{\scriptstyle\prime}$\fi}               
\def\parcs{\saa=.07em \sbb=.03em
     \ifmmode \hbox{\rlap{.}}^{\scriptstyle\prime\kern -\sbb\prime}\hbox{\kern -\saa}
     \else \rlap{.}$^{\scriptstyle\prime\kern -\sbb\prime}$\kern -\saa\fi}

\shortauthors{Murgia et al.}

\begin{document}
\makebox[17cm][r]{\textcopyright 2026 all rights reserved.}

\title{\spherex\ Mapping of Diffuse PAH and \HII\ Emission in the Galactic Plane}

\author[0009-0002-0149-9328]{Giulia~Murgia}%
\affiliation{Department of Physics, California Institute of Technology, 1200 E. California Boulevard, Pasadena, CA 91125, USA}%
\email[show]{gmurgia@caltech.edu}%
\author[0000-0002-7471-719X]{Ari~J.~Cukierman}%
\affiliation{Department of Physics, California Institute of Technology, 1200 E. California Boulevard, Pasadena, CA 91125, USA}%
\email{ajcukier@caltech.edu}%
\author[0000-0001-7449-4638]{Brandon~S.~Hensley}%
\affiliation{Jet Propulsion Laboratory, California Institute of Technology, 4800 Oak Grove Drive, Pasadena, CA 91109, USA}%
\email{brandon.s.hensley@jpl.nasa.gov}%
\author[0000-0002-3993-0745]{Matthew~L.~N.~Ashby}%
\affiliation{Center for Astrophysics $|$ Harvard \& Smithsonian, Optical and Infrared Astronomy Division, Cambridge, MA 01238, USA}%
\email{mashby@cfa.harvard.edu}%
\author[0000-0002-5710-5212]{James~J.~Bock}%
\affiliation{Department of Physics, California Institute of Technology, 1200 E. California Boulevard, Pasadena, CA 91125, USA}%
\affiliation{Jet Propulsion Laboratory, California Institute of Technology, 4800 Oak Grove Drive, Pasadena, CA 91109, USA}%
\email{jjb@astro.caltech.edu}%
\author[0000-0001-5929-4187]{Tzu-Ching~Chang}%
\affiliation{Jet Propulsion Laboratory, California Institute of Technology, 4800 Oak Grove Drive, Pasadena, CA 91109, USA}%
\affiliation{Department of Physics, California Institute of Technology, 1200 E. California Boulevard, Pasadena, CA 91125, USA}%
\email{tzu@caltech.edu}%
\author[0009-0000-3415-2203]{Shuang-Shuang~Chen}%
\affiliation{Department of Physics, California Institute of Technology, 1200 E. California Boulevard, Pasadena, CA 91125, USA}%
\email{schen6@caltech.edu}%
\author[0000-0002-5437-0504]{Yun-Ting~Cheng}%
\affiliation{Department of Physics, California Institute of Technology, 1200 E. California Boulevard, Pasadena, CA 91125, USA}%
\affiliation{Jet Propulsion Laboratory, California Institute of Technology, 4800 Oak Grove Drive, Pasadena, CA 91109, USA}%
\email{ycheng3@caltech.edu}%
\author[0000-0001-6320-261X]{Yi-Kuan~Chiang}%
\affiliation{Academia Sinica Institute of Astronomy and Astrophysics (ASIAA), No. 1, Section 4, Roosevelt Road, Taipei 10617, Taiwan}%
\email{ykchiang@asiaa.sinica.edu.tw}%
\author[0000-0002-3892-0190]{Asantha~Cooray}%
\affiliation{Department of Physics \& Astronomy, University of California Irvine, Irvine, CA 92697, USA}%
\email{acooray@uci.edu}%
\author[0000-0002-4650-8518]{Brendan~P.~Crill}%
\affiliation{Jet Propulsion Laboratory, California Institute of Technology, 4800 Oak Grove Drive, Pasadena, CA 91109, USA}%
\affiliation{Department of Physics, California Institute of Technology, 1200 E. California Boulevard, Pasadena, CA 91125, USA}%
\email{bcrill@jpl.nasa.gov}%
\author[0000-0001-7432-2932]{Olivier~Dor\'{e}}%
\affiliation{Jet Propulsion Laboratory, California Institute of Technology, 4800 Oak Grove Drive, Pasadena, CA 91109, USA}%
\affiliation{Department of Physics, California Institute of Technology, 1200 E. California Boulevard, Pasadena, CA 91125, USA}%
\email{olivier.dore@caltech.edu }%
\author[0009-0002-0098-6183]{C.~Darren~Dowell}%
\affiliation{Jet Propulsion Laboratory, California Institute of Technology, 4800 Oak Grove Drive, Pasadena, CA 91109, USA}%
\affiliation{Department of Physics, California Institute of Technology, 1200 E. California Boulevard, Pasadena, CA 91125, USA}%
\email{charles.d.dowell@jpl.nasa.gov}%
\author[0000-0002-9382-9832]{Andreas~L.~Faisst}%
\affiliation{IPAC, California Institute of Technology, MC 100-22, 1200 E California Blvd, Pasadena, CA 91125, USA}%
\email{afaisst@caltech.edu}%
\author[0000-0002-5599-4650]{Joseph~L.~Hora}%
\affiliation{Center for Astrophysics $|$ Harvard \& Smithsonian, Optical and Infrared Astronomy Division, Cambridge, MA 01238, USA}%
\email{jhora@cfa.harvard.edu}%
\author[0000-0001-5812-1903]{Howard~Hui}%
\affiliation{Department of Physics, California Institute of Technology, 1200 E. California Boulevard, Pasadena, CA 91125, USA}%
\affiliation{Jet Propulsion Laboratory, California Institute of Technology, 4800 Oak Grove Drive, Pasadena, CA 91109, USA}%
\email{hhui@caltech.edu}%
\author[0000-0002-5016-050X]{Miju~Kang}%
\affiliation{Korea Astronomy and Space Science Institute (KASI), 776 Daedeok-daero, Yuseong-gu, Daejeon 34055, Republic of Korea}%
\email{mjkang@kasi.re.kr}%
\author[0000-0002-3470-2954]{Jae~Hwan~Kang}%
\affiliation{Department of Physics, California Institute of Technology, 1200 E. California Boulevard, Pasadena, CA 91125, USA}%
\email{jkang7@caltech.edu}%
\author[0009-0003-8869-3365]{Phil~M.~Korngut}%
\affiliation{Department of Physics, California Institute of Technology, 1200 E. California Boulevard, Pasadena, CA 91125, USA}%
\email{pkorngut@caltech.edu}%
\author[0000-0002-3455-1826]{Dennis~Lee}%
\affiliation{Jet Propulsion Laboratory, California Institute of Technology, 4800 Oak Grove Drive, Pasadena, CA 91109, USA}%
\email{dennisl@jpl.nasa.gov}%
\author[0000-0003-3119-2087]{Jeong-Eun Lee}%
\affiliation{Department of Physics and Astronomy, Seoul National University, 1 Gwanak-ro, Gwanak-gu, Seoul 08826, Republic of Korea}%
\email{lee.jeongeun@snu.ac.kr}%
\author[0000-0003-1954-5046]{Bomee~Lee}%
\affiliation{Korea Astronomy and Space Science Institute (KASI), 776 Daedeok-daero, Yuseong-gu, Daejeon 34055, Republic of Korea}%
\affiliation{IPAC, California Institute of Technology, MC 100-22, 1200 E California Blvd, Pasadena, CA 91125, USA}%
\email{bomee@kasi.re.kr}%
\author[0000-0002-9548-1526]{Carey~M.~Lisse}%
\affiliation{Johns Hopkins University, 3400 N Charles St, Baltimore, MD 21218, USA}%
\affiliation{Johns Hopkins University Applied Physics Laboratory, Laurel, MD 20723, USA}%
\email{carey.lisse@jhuapl.edu}%
\author[0000-0001-5382-6138]{Daniel~C.~Masters}%
\affiliation{IPAC, California Institute of Technology, MC 100-22, 1200 E California Blvd, Pasadena, CA 91125, USA}%
\email{dmasters@ipac.caltech.edu}%
\author[0000-0002-6025-0680]{Gary~J.~Melnick}%
\affiliation{Center for Astrophysics $|$ Harvard \& Smithsonian, Optical and Infrared Astronomy Division, Cambridge, MA 01238, USA}%
\email{gmelnick@cfa.harvard.edu}%
\author[0000-0003-3393-2819]{Mary~H.~Minasyan}%
\affiliation{Department of Physics, California Institute of Technology, 1200 E. California Boulevard, Pasadena, CA 91125, USA}%
\email{minasyan@caltech.edu}%
\author[0000-0001-9368-3186]{Chi~H.~Nguyen}%
\affiliation{Department of Physics, California Institute of Technology, 1200 E. California Boulevard, Pasadena, CA 91125, USA}%
\email{chnguyen@caltech.edu}%
\author[0000-0002-5158-243X]{Roberta~Paladini}%
\affiliation{IPAC, California Institute of Technology, MC 100-22, 1200 E California Blvd, Pasadena, CA 91125, USA}%
\email{paladini@ipac.caltech.edu}%
\author[0000-0003-1841-2241]{Volker~Tolls}%
\affiliation{Center for Astrophysics $|$ Harvard \& Smithsonian, Optical and Infrared Astronomy Division, Cambridge, MA 01238, USA}%
\email{vtolls@cfa.harvard.edu}%
\author[0000-0001-7254-1285]{Robin~Y.~Wen}%
\affiliation{Department of Physics, California Institute of Technology, 1200 E. California Boulevard, Pasadena, CA 91125, USA}%
\email{ywen@caltech.edu}%
\author[0000-0003-4990-189X]{Michael~W.~Werner}%
\affiliation{Jet Propulsion Laboratory, California Institute of Technology, 4800 Oak Grove Drive, Pasadena, CA 91109, USA}%
\email{michael.w.werner@jpl.nasa.gov}%
\author[0000-0001-8253-1451]{Michael~Zemcov}%
\affiliation{School of Physics and Astronomy, Rochester Institute of Technology, 1 Lomb Memorial Dr., Rochester, NY 14623, USA}%
\affiliation{Jet Propulsion Laboratory, California Institute of Technology, 4800 Oak Grove Drive, Pasadena, CA 91109, USA}%
\email{mbzsps@rit.edu}

\begin{abstract}
We present preliminary \spherex\ maps of diffuse Galactic emission tracing polycyclic aromatic hydrocarbons~(PAHs) and ionized hydrogen gas, and we study their relationship across the Galactic plane. Since its launch in early 2025, the \spherex\ space telescope has been conducting an all-sky near-infrared spectral survey from~$\wlmin$ to~$\wlmax~\um$. We produce a large-scale map of the $3.3$-$\um$~PAH emission feature, which is bright and detectable throughout the Galactic plane, and find a strong correlation with the thermal dust radiance measured by \planck. We also trace ionized hydrogen gas by producing a map of Brackett-$\alpha$ emission at $4.05~\um$. By combining the two maps, we identify extended shells of PAH~emission associated with photodissociation regions surrounding ionized gas. We construct a PAH abundance map and find a significant anticorrelation between PAH abundance and ionized hydrogen, indicating systematic PAH depletion within ionized gas regions across the Galactic plane and demonstrating that ionizing radiation is a dominant driver of PAH abundance variations. These early \spherex\ results provide a large-scale view of~PAHs and ionized hydrogen and preview the capability of the mission to map diffuse emission in the interstellar medium.
\end{abstract}


\section{Introduction}
\label{sec:intro}
\setcounter{footnote}{0}

The Spectro-Photometer for the History of the Universe, Epoch of Reionization and Ices Explorer~\citep[SPHEREx,][]{Bock2026} is a NASA Medium-class Explorer~(MIDEX) mission launched in March 2025 that is performing an all-sky near-infrared spectral survey from~$\wlmin$ to~$\wlmax~\um$ with a resolving power of $R \sim 35$--$130$. The mission operates from low Earth orbit and uses a telescope with a $3.5^\circ \times 11^\circ$~field of view and a pixel size of~6\parcs{15}, equipped with six linear variable filters~(LVFs) for spectroscopic sensitivity~\citep{Korngut2026}. Within each exposure image, the central wavelength varies across the field of view~\citep{Hui2026}. Multiple exposures sample each line of sight across 102 nominal spectral channels. The mission surveys the entire sky approximately twice per year, producing four full-sky surveys over its baseline lifetime.

Beyond its core scientific goals~\citep{Bock2026}, \spherex's continuous all-sky spectral coverage enables a new view of the interstellar medium~(ISM). Between~$\wlmin$ and~$\wlmax~\um$, \spherex\ captures several spectral features, including hydrogen recombination lines and the $3.3$-$\um$~emission feature associated with polycyclic aromatic hydrocarbons~(PAHs). These features trace complementary components of the ISM and allow us to probe the distribution of dust, gas, and radiation across the entire Galaxy. \citet{Hora2026} presented wide-field spectral maps of the Cygnus-X and the North America Nebula regions, showcasing \spherex's ability to trace interstellar ices and PAH emission over large areas. 

In this paper, we present maps of diffuse PAHs and ionized hydrogen~(\HII) emission over most of the  Galactic plane, and study PAH abundance and depletion in relation to the distribution of ionized gas. A key advantage of \spherex\ for ISM studies of this kind is its spectroscopic coverage, which allows individual emission features to be isolated rather than blended within broad photometric bands. Previous Galactic surveys with instruments such as the Wide-Field Infrared Survey Explorer~\citep[WISE,][]{Wright2010} and the Spitzer Space Telescope~\citep{Werner2004} used broad photometric bands, making it difficult to separate the specific physical components contributing to the observed signal. In contrast, the narrow spectral channels of \spherex\ enable direct mapping of the $3.3$-$\um$~PAH feature, as well as individual hydrogen recombination lines.

Well mixed with interstellar dust, PAHs are molecules made of carbon and hydrogen atoms arranged in multiply-connected six-member ring structures. They efficiently absorb near-ultraviolet photons and re-emit this energy in a series of mid-infrared bands. The emission features are produced by the stretching and bending modes of the aromatic C-H and C-C bonds~\citep{Allamandola1985, Tielens2008}, with prominent bands at 3.3, 6.2, 7.7, 8.6, 11.3, and $12.7~\um$~\citep{LegerPuget1984}. In the Galactic plane, PAHs are widespread and appear in the diffuse ISM, in molecular clouds, and especially in the shells surrounding \HII~regions~\citep{Smith2007, Galliano2008, PereiraSantaella2010}. Young, massive stars produce strong ultraviolet radiation that ionizes hydrogen, creating \HII~regions. Inside these regions, PAHs are often destroyed by the harsh radiation~\citep{Sloan1997, Mori2014}. Just outside, in photodissociation regions~(PDRs), the non-ionizing ultraviolet radiation is still strong enough to excite these molecules, but not sufficiently intense to destroy them~\citep{WeingartnerDraine2001}. This produces the characteristic PAH shells observed around many \HII~regions~\citep{Watson2009, Watson2010, Winston2012}. 

Measuring how PAH emission changes from the interior of an \HII~region to its surrounding PDR provides a direct way to study PAH depletion and the impact of stellar radiation on these molecules. Because the $3.3$-$\um$~feature is most sensitive to small, neutral PAHs~\citep{DraineLi2007, Rigopoulou2021}, mapping this band provides a way to identify where the smallest grains survive and how they are processed as the radiation field changes. Previous studies in the Milky Way have largely focused on individual clouds or targeted fields using infrared imaging from instruments such as Spitzer~\citep[e.g.,][]{Churchwell2007,Paradis2011}. With its all-sky spectral coverage, \spherex\ now enables the study of the distribution of $3.3$-$\um$~PAH emission across the entire Milky Way.

Complementary extragalactic studies have investigated this effect in external galaxies. Extragalactic Spitzer observations have shown that PAH abundance is spatially variable and strongly suppressed within \HII~regions, as demonstrated in the Small Magellanic Cloud by \citet{Sandstrom2010}. More recently, JWST and MUSE observations from the PHANGS project showed a strong anticorrelation between PAH fraction and ionization parameter across $\sim$1500~\HII~regions in four nearby star-forming galaxies~\citep{Egorov2023}. This result has since been extended to 42~nearby galaxies~\citep{Egorov2025}, establishing PAH depletion near ionized gas as a common feature of star-forming environments. 

At the same time, \spherex\ spectral coverage enables mapping of multiple hydrogen recombination lines, which trace warm ionized gas~\citep{Garay1999, Moises2011}. To limit the scale of computation in this initial study, we restrict the analysis to a single line, \bralpha\ at $4.05~\um$. The \bralpha\ line arises from the $n=5 \to 4$ transition and traces ionized gas produced by ultraviolet photons from massive OB~stars~\citep{Yano2016}. This recombination line is among the brightest in the \spherex\ wavelength range. At $4.05~\um$, \spherex\ offers significantly higher spectral resolution than at Paschen~$\alpha$~(\paalpha, $1.87~\um$), with a resolving power of $R \approx 110$ at~$4.05~\um$ compared to $R \approx 41$ at $1.87~\um$. Its longer wavelength also makes \bralpha\ significantly less affected by dust extinction than shorter-wavelength lines such as \paalpha. In addition, the zodiacal foreground reaches a minimum near~$\sim 3.5~\um$ in the \spherex\ survey~\citep{Bock2026}, placing \bralpha\ close to the wavelength of minimum zodiacal sky brightness and resulting in lower zodiacal contamination than Pfund~$\beta$~(\pfbeta, $4.65~\um$). These factors make \bralpha\ a particularly clean and robust tracer of ionized gas for this initial investigation.

Our maps show large-scale diffuse structures across the Galactic plane and contain many instances of PAH~shells surrounding ionized gas. By comparing PAH emission with dust radiance traced by the \planck\ mission, we build a PAH abundance map that highlights where PAHs survive or are depleted. These maps allow us to examine the relationships among PAH abundance, ionized gas, and dust across the Galactic plane and to quantify large-scale depletion. Importantly, both the PAH emission and the ionized-gas tracer used in this work are derived from the same \spherex\ observations and processed in a fully self-consistent manner.

Although \spherex\ will ultimately produce full-sky maps, the correlation analyses in this work are restricted to maps with latitude cuts and masking designed to isolate strong diffuse emission only near the Galactic plane. This study should therefore be viewed as an initial step toward full-sky diffuse-emission maps.

The structure of this paper is as follows. In \secref{sec:methods}, we describe the image-processing and map-making methods to produce maps of large-scale diffuse structures from \spherex\ observations. In \secref{sec:maps}, we present the resulting diffuse-emission maps of the $3.3$-$\um$~PAH feature and \bralpha\ emission. In \secref{sec:results}, we analyze pixel-by-pixel correlations among PAH emission, dust radiance, and ionized gas. Finally, we discuss future investigations in \secref{sec:discussion} and summarize our results in \secref{sec:conclusion}.

\section{Image processing and map making}
\label{sec:methods}

In this section, we describe the image-processing and map-making steps used to create diffuse emission maps from calibrated \spherex\ spectral images. Our analysis begins with the Level-2~(L2) calibrated spectral images produced by the \spherex\ Science Data Center~(SSDC), which are publicly available on the IRSA website.\footnote{\url{https://irsa.ipac.caltech.edu/Missions/spherex.html}} The \spherex\ image processing pipeline described in \cite{Akeson2025} has already applied the standard detector and optical calibrations and provides images with solved astrometry. In this work, we use the Quick Release 2~(QR2) L2 data products available as of December 15, 2025~\citep{SPHERExQR2}.

A subset of \spherex\ exposures taken toward very dense stellar fields currently lack reliable astrometric solutions due to source crowding~\citep{Akeson2025}. We exclude exposures with an astrometric rms $> 1\arcsec$, which accounts for approximately 2\% of the total available exposures, concentrated primarily toward the Galactic center. Although these exposures are available and can be reprojected onto the map grid, the astrometric uncertainty prevents the application of the source-masking and diffuse-emission extraction steps described below. Therefore, these exposures are excluded from the analysis presented in this paper. We note that excluding the Galactic center does not limit the scope of this work. The central regions of the Galaxy are characterized by extremely high stellar densities and complex environments, making the separation of diffuse emission from unresolved stellar light particularly challenging. Therefore, the absence of these regions in the diffuse-emission maps is appropriate for the goals of this study. The astrometric solutions in these affected regions are expected to improve in future data releases.

\subsection{Image Processing}
\label{subsec:image}

Through the action of its associated LVF, each image detects light over a range of wavelengths. We subdivide each L2 spectral image into 17 ``channels" that correspond to wavelength intervals. The channels appear as horizontal stripes across the exposure, with a slight upward curvature~(``smile") across the field arising from the LVF wavelength response. There are six detector arrays, referred to as Bands~1--6, giving a total of 102 channels spanning~$\wlmin$--$\wlmax~\um$. The $3.3$-$\um$~PAH feature lies within Band~4~($2.38$--$3.87~\um$), while \bralpha\ at~$4.05~\um$ falls within Band~5~($3.79$--$4.44~\um$).

\begin{figure*} 
\centering
\includegraphics[width=1\textwidth]{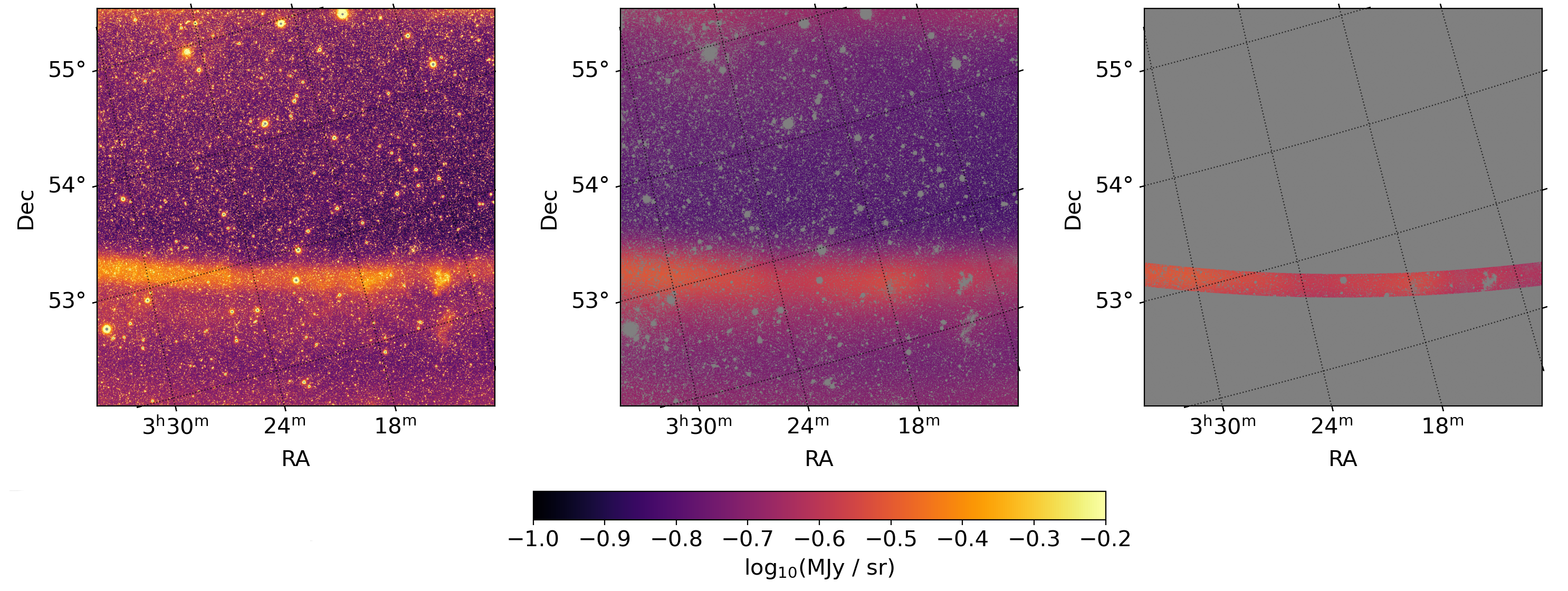}
\caption{Example Band-4 image illustrating the processing steps used to extract diffuse emission. Left:~L2 calibrated spectral image dominated by stars and galaxies. Wavelength varies continuously from~$2.42~\um$~at the top to~$3.82~\um$~at the bottom due to the LVF response. The bright, nearly horizontal band corresponds to enhanced emission in the wavelength range containing the $3.3$-$\um$~PAH feature. Middle: the same exposure after source masking, sigma clipping, flagging, and smoothing, revealing diffuse emission. Right: an extracted wavelength range centered on the $3.3$-$\um$~PAH feature~($\lambda_{\rm min} = 3.26~\um$ to $\lambda_{\rm max} = 3.36~\um$), used to construct the PAH emission map. Different wavelength ranges can be extracted to produce maps of other spectral features.}
\label{fig:exposures}
\end{figure*}

The left panel in \figref{fig:exposures} shows an example Band-4 image, covering wavelengths from 2.38 to $3.87~\um$. Bright stars and galaxies dominate the field. Diffuse emission is also present, but it is hidden by the point sources. One of the spectral stripes appears brighter than the others: this corresponds to the $3.3$-$\um$~PAH feature, which produces enhanced emission in that portion of the spectrum. In the steps that follow, we apply additional processing to these L2 images using the map-making methodologies described in \cite{Cukierman2026}.

In order to identify and mask discrete sources, such as bright stars and galaxies, and reveal the otherwise obscured diffuse emission, we apply a simulation-based threshold mask. Using the \spherex\ Sky \skysim\ described in \cite{Crill2025}, we generate simulated images of stars and galaxies from sources in the \spherex\ Reference Catalog ~\citep{Akeson2025, Yang2026}. We define a surface-brightness threshold and mask all pixels where the simulated flux exceeds the chosen threshold. In this work, we mask galaxies and stars to magnitude $m_{\rm AB} < 16$, where the magnitude is evaluated at the central wavelength of each band~($3.05~\um$ for Band~4 and $4.10~\um$ for Band~5), using an intensity threshold of $0.2~\MJysr$. The adopted values represent a compromise between removing bright sources and retaining sky coverage. We find that our maps are only weakly sensitive to these masking parameters.

To further clean each exposure of outlier pixels, we apply the L2 pixel-quality flags~\citep{Akeson2025}, which identify detector artifacts such as transients and non-functional pixels. We then perform sigma clipping, which removes pixels whose values deviate by more than two standard deviations from the local background level, computed independently within each wavelength bin~\citep{Cukierman2026}. The total fraction of masked pixels varies among exposures. For the example shown in \figref{fig:exposures}, 28.46\% of pixels are masked after applying the source mask, quality flags, and sigma clipping. Finally, we smooth the images to $20\arcmin$ to suppress small-scale features. This scale is several times larger than the map pixel size, so the smoothing acts as an anti-aliasing filter and also helps to reduce LVF gradients. The middle panel in \figref{fig:exposures} shows the same exposure after source masking, flagging, sigma clipping, and smoothing. Discrete sources and small-scale features are removed, isolating diffuse emission that varies across the field. 

Because the effective wavelength varies continuously across each \spherex\ image, we select the wavelength range corresponding to the spectral feature of interest before constructing maps. The right panel in \figref{fig:exposures} illustrates an example extraction for one of the 102 channels, using the wavelength interval~$\lambda_{\rm min} = 3.26~\um$ to $\lambda_{\rm max} = 3.36~\um$. This step makes it possible to extract specific spectral features, such as the $3.3$-$\um$~PAH feature in this case. Each extracted channel is then passed to the reprojection algorithm described in \secref{subsec:mapmaking}.

\subsection{Map-making}
\label{subsec:mapmaking}

To construct maps from \spherex\ observations, we define a target sky grid and reproject individual exposures onto it. Each pixel of each \spherex\ spectral image contributes surface-brightness measurements to the sky pixel onto which it projects, and the contributions from multiple pixels are averaged. As more images are incorporated, coverage expands, and noise averages down~(see~\citet{Cukierman2026} for more details).

\subsubsection{\healpix\ Representation}

Full-sky maps are stored using a \healpix\footnote{HEALPix: Hierarchical Equal Area isoLatitude Pixelization, \url{http://healpix.sourceforge.net}.} grid, which divides the celestial sphere into equal-area pixels~\citep{Gorski2005}. The resolution is controlled by the parameter $\nside$, with the total number of pixels given by $N_{\rm pix} = 12 \nside^2.$ Higher $\nside$ corresponds to finer angular resolution and a larger number of pixels. Each pixel has a solid angle $\Omega_{\rm pix} = 4\pi / N_{\rm pix} = 4\pi / (12 \nside^2)$. A characteristic angular scale for a pixel can be estimated as $\theta_{\rm pix} \approx \sqrt{\Omega_{\rm pix}}$. For this work, we adopt $\nside = 512$, which corresponds to $\theta_{\rm pix} \approx 7\arcmin$.

While the analyses in this work focus on maps with latitude cuts that isolate strong diffuse emission near the Galactic plane, the use of the \healpix\ framework ensures that the methods developed here can be extended to the full sky in future releases.

\subsubsection{Map Reprojection}

We reproject one exposure and one spectral channel at a time and reprojected images are coadded in map space. Image pixels are binned into map pixels with an intensity weighting~\citep{Cukierman2026}. We also accumulate a hit-count map, defined as the number of \spherex\ pixels contributing to each map pixel. 

To validate the map-making pipeline, we apply our reprojection algorithm to simulated \spherex\ data~\citep{Crill2025}. The \skysim\ produces spectral images and catalogs for individual components, including Diffuse Galactic Light~(DGL), zodiacal light, photon noise, stars, galaxies, read noise, and dark current. We use these simulated maps to verify that the reprojection recovers the input sky structure, and test source masking and smoothing steps to ensure that diffuse features are revealed.

\subsubsection{Full-sky Spectral Cube}

\begin{figure*} 
    \centering
    \includegraphics[width=1\textwidth]{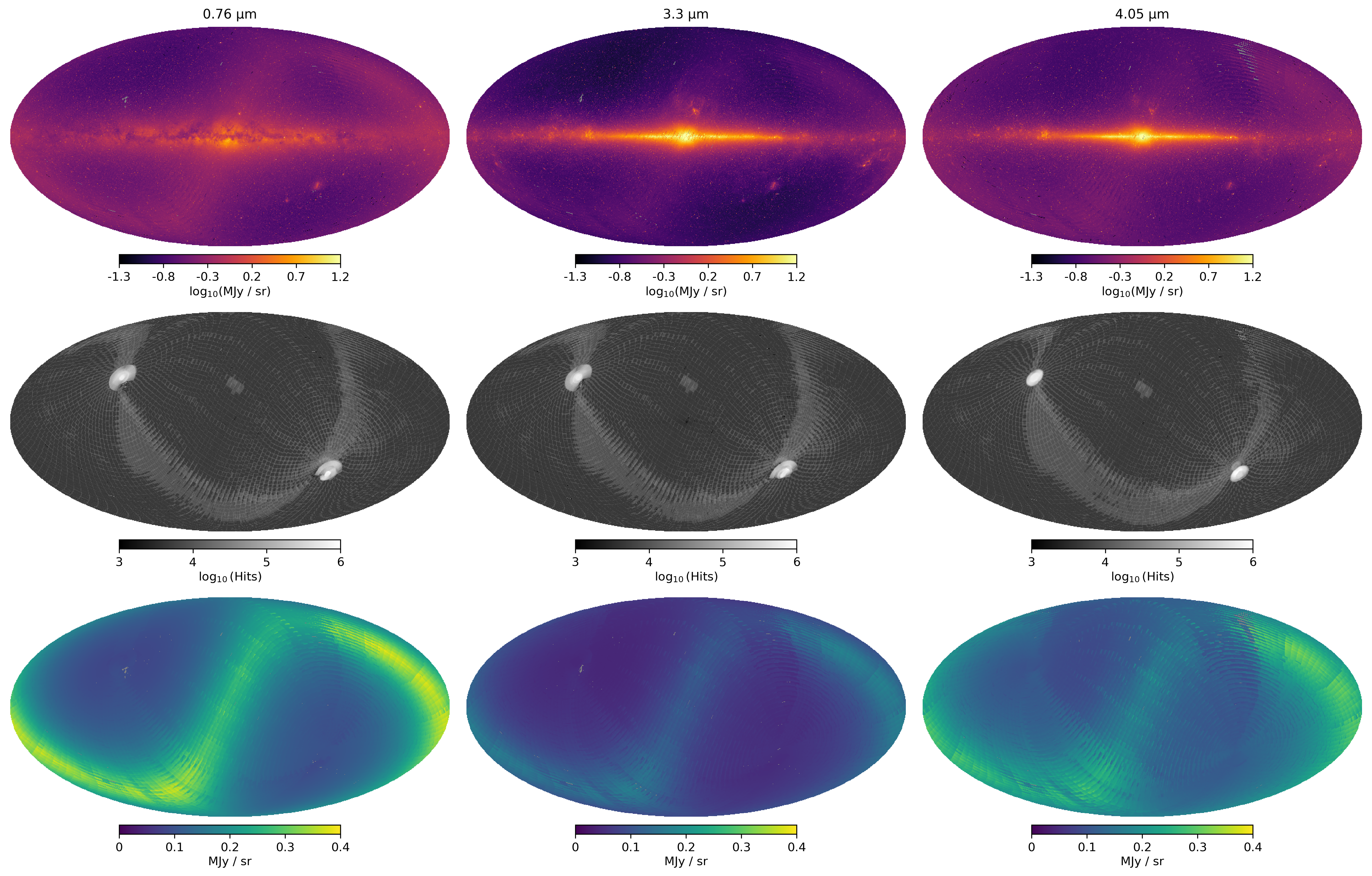}
    \caption{Full-sky maps for three representative wavelengths, shown in Mollweide projection in Galactic coordinates. Top row:~full-sky maps at $0.76~\um$, $3.3~\um$ near the PAH feature, and $4.05~\um$ near \bralpha, produced by directly extracting and reprojecting individual spectral channels without source masking. In addition to stellar emission, a band of zodiacal light running from the lower left to the upper right of each map is visible, especially at $0.76~\um$. It shows discrete discontinuities arising from time variability of the observed zodiacal light. 
    Middle row:~corresponding hit-count maps indicating the number of contributing exposures per pixel.
    Bottom row:~simulated zodiacal-light full-sky maps processed through the same pipeline. The similarity between the simulated and observed patterns highlights the contribution of zodiacal emission to the real maps.}
    \label{fig:fullskymap}
\end{figure*}

To illustrate the importance of source masking, we first consider the case in which the images are smoothed only. The resulting maps at 0.76, 3.3, and $4.05~\um$ are shown in the top row of \figref{fig:fullskymap}. 

These full-sky maps are created by reprojecting all \spherex\ exposures, including those with poor astrometric solutions~(as no source masking or small-scale measurements are performed at this stage, and the maps are used only to illustrate the sky coverage and scanning strategy of the survey). Because no masking or filtering has been applied, stars and galaxies remain significant in these maps. By repeating this procedure for each of the 102 nominal spectral channels, we can produce a set of full-sky maps that can be viewed as an all-sky spectral cube.\footnote{\url{https://www.jpl.nasa.gov/images/pia26600-spherexs-first-all-sky-map/}} Exposures near the Galactic center, which are affected by poor astrometric solutions, are excluded from subsequent diffuse-emission processing and scientific analysis.

The second row of \figref{fig:fullskymap} shows the corresponding hit-count maps. The hit-count map reflects the number of image pixels contributing to each map pixel and therefore shows the survey’s coverage pattern~\citep{Bryan2025}. The two deep fields at the north and south ecliptic poles are clearly visible as regions with very high hit counts.

\subsubsection{Zodiacal-light Considerations}
\label{subsec:zodi}

In addition to the stellar contamination, a diffuse glow is visible along the ecliptic plane. This is zodiacal light, produced by a combination of scattered sunlight and thermal emission from dust grains in the Solar System. The third row of \figref{fig:fullskymap} shows examples zodiacal light simulation processed through the same reprojection pipeline as the real data for the same subset of wavelengths. The zodiacal emission is modeled using a modified version of the DIRBE-based interplanetary dust model of \citet{Kelsall1998}, as described in \citet{Crill2025}. These maps show a gradient perpendicular to the ecliptic plane with discrete discontinuities arising from time variability in the zodiacal light combined with the survey scan strategy. \spherex\ scans the sky in repeated passes, returning to the same regions at different times to fill in coverage. The zodiacal foreground is time and location dependent and therefore changes between these visits causing exposures of the same sky region taken at different times to have different zodiacal light levels. When these data are combined into full-sky maps, the varying zodiacal light levels appear as discontinuities aligned with the scan pattern.

When we compare the zodiacal light signal to the real \spherex\ data at low Galactic latitudes, the diffuse Galactic emission exceeds the zodiacal contribution by more than an order of magnitude in the $3.3~\um$ and $4.05~\um$ channels used in this work. For simplicity in this initial analysis, we restrict ourselves to the regions defined by the \planck\ 70\% Galactic mask.\footnote{HFI\_Mask\_GalPlane-apo0\_2048\_R2.00.fits, available from the Planck Legacy Archive (\url{https://pla.esac.esa.int})} This is a standard \planck\ mask derived from the 353~GHz dust map and constructed to retain 70\% of the sky by excluding the brightest thermal dust emission near the Galactic plane. In this work, we instead select the complementary sky area, focusing on the bright Galactic plane emission in 30\% of the sky. In this region, the Galactic emission is sufficiently bright that the zodiacal foreground is subdominant. Therefore, no zodiacal subtraction is applied.

\subsubsection{Spectral Interpolation}
\label{subsec:specint}

When mapping sharp spectral features, the wavelength spread within a nominal channel due to the LVF response can produce scan-aligned gradients in the maps. The smoothing applied to each exposure partially mitigates this effect. To further correct for these gradients, we can apply spectral interpolation at the map level. We subdivide each nominal spectral channel into a set of narrowband sub-channels, each covering a smaller wavelength range and therefore more closely approximating a single wavelength. We then interpolate along the spectral axis for each map pixel to evaluate the intensity at a fixed target wavelength. Further details and validation of this method are presented in~\cite{Cukierman2026}.

\begin{figure*} 
\centering
\begin{subfigure}{\textwidth}
  \centering
  \includegraphics[width=\linewidth]{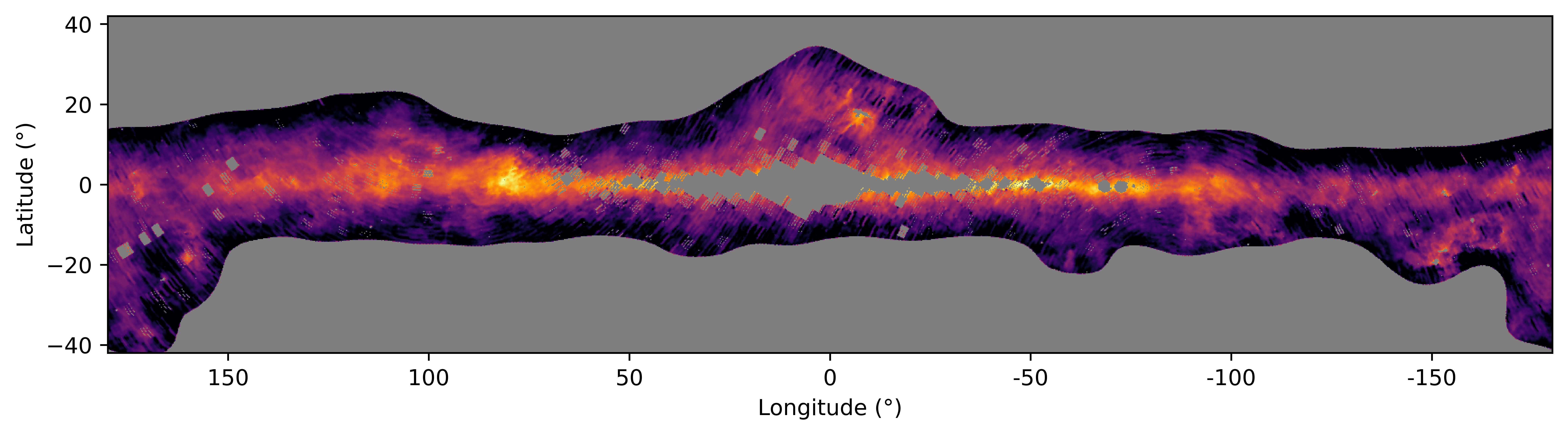}
\end{subfigure}
\vspace{1em}
\begin{subfigure}{\textwidth}
  \centering
  \includegraphics[width=\linewidth]{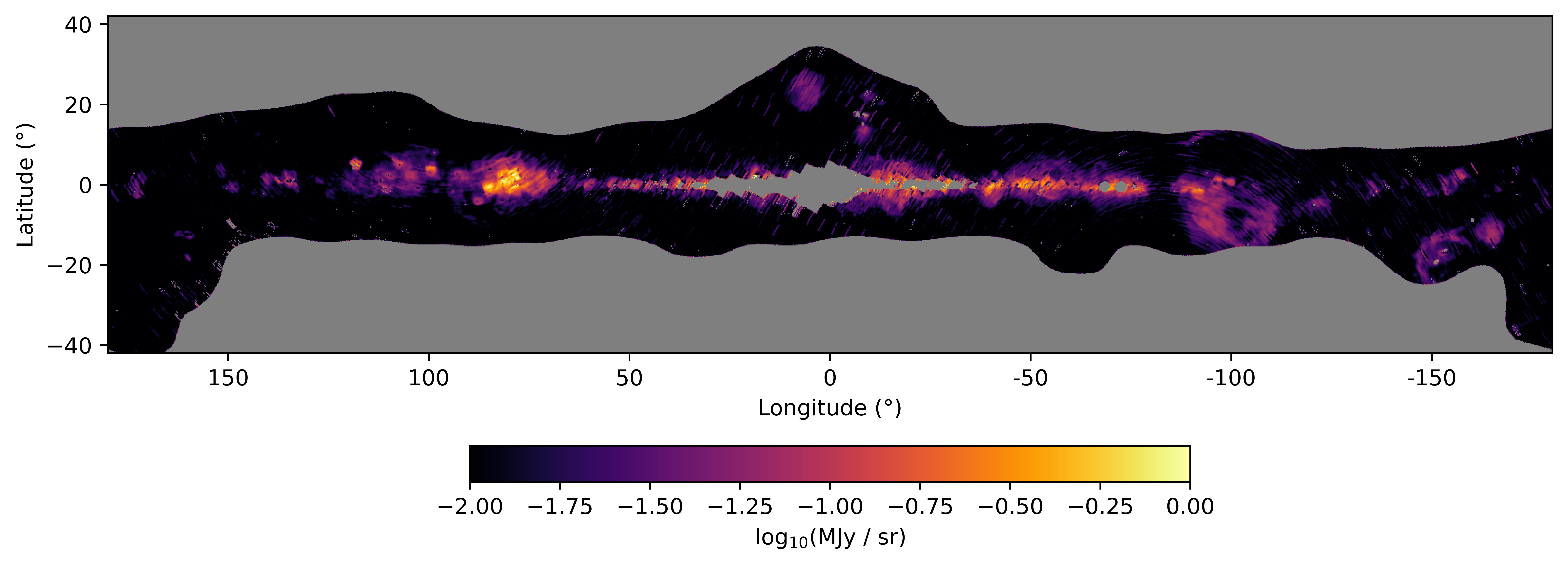}
\end{subfigure}
\caption{Maps of diffuse emission along the Galactic plane after continuum subtraction and application of the \planck\ 70\% Galactic mask. The gap near the Galactic center reflects exposures excluded due to large astrometric uncertainties. Improved astrometric solutions in future data releases are expected to make this region accessible for this type of analysis. Top: Map of $3.3$-$\um$~PAH emission. Bottom: Map of $4.05$-$\um$~\bralpha\ emission tracing ionized gas. Some stripes are visible as artifacts from the scanning pattern and time-variable foregrounds.}
\label{fig:line_maps}
\end{figure*}

\begin{figure*} 
    \centering
    \includegraphics[width=1\textwidth]{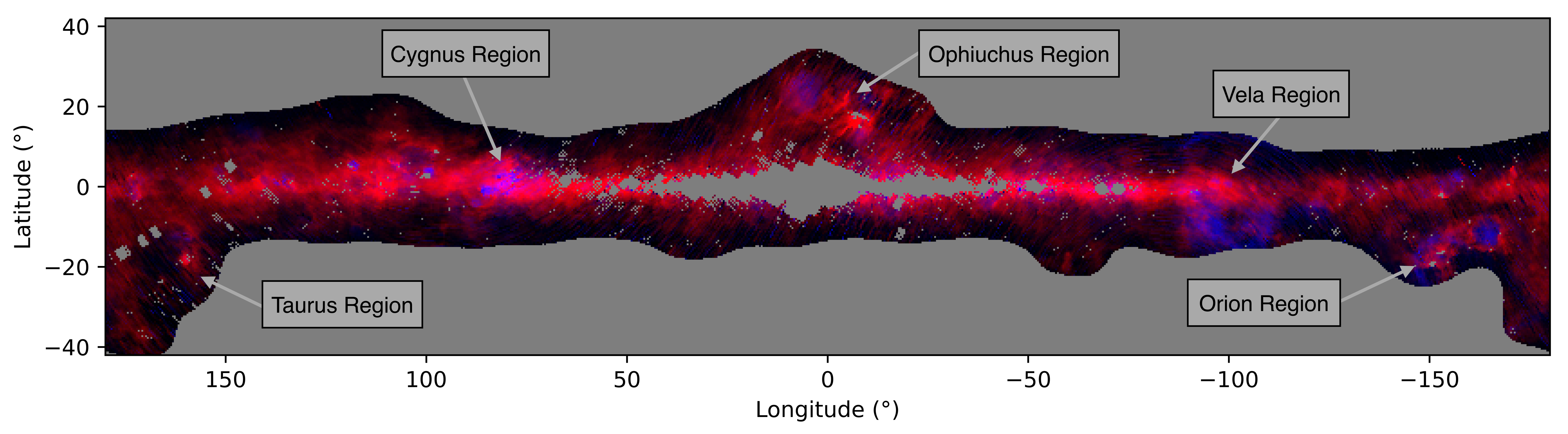}
    \caption{Two-color composite map of the Galactic plane. Red shows $3.3$-$\um$~PAH emission, and blue shows $4.05$-$\um$~\bralpha. This rendering highlights the spatial relationship between the diffuse components with PAH emission outlining shells around regions of ionized gas. Labels indicate prominent star-forming regions across the Galactic plane.
    }
    \label{fig:rgb}
\end{figure*}

\section{Maps of Diffuse Galactic Emission}
\label{sec:maps}

We apply the image-processing and map-making pipeline to construct diffuse-emission maps for the spectral features of interest. For each exposure, we apply pixel masking and smoothing and reproject to a \healpix\ grid with $\nside=512$. We adopt this resolution in combination with the $20\arcmin$ smoothing applied to each exposure to emphasize large-scale, diffuse features and help to reduce LVF gradients. This choice sacrifices fine angular structure but provides a practical compromise between resolution, map uniformity, and computational cost for this initial analysis.

For the PAH feature, we extract emission from Band~4 by applying the spectral interpolation described in \secref{subsec:specint} to evaluate the intensity at a target wavelength of $3.3~\um$. Similarly, the \bralpha\ feature is extracted from Band~5 by interpolating to a target wavelength of $4.05~\um$.

Moreover, we perform continuum subtraction at the map level. For each spectral feature, we fit and subtract a linear continuum using two spectrally interpolated maps at neighboring wavelengths. For the $3.3$-$\um$~PAH feature, we use maps interpolated to 3.15 and $3.6~\um$, while for \bralpha\ at $4.05~\um$ we use maps interpolated to 3.9 and $4.2~\um$. We note that this continuum subtraction also partially mitigates zodiacal light contamination, since zodiacal emission is spectrally smooth in this wavelength range and is therefore absorbed by the linear continuum fit.

\subsection{Diffuse PAH and \bralpha\ Emission Maps}
\label{ssec:galacticmaps}

\figref{fig:line_maps} shows the resulting $3.3$-$\um$~PAH map on the top panel and the \bralpha\ map at $4.05~\um$ on the bottom panel. The PAH map provides a large-scale view of $3.3$-$\um$~PAH emission, which is bright and detectable throughout the Galactic plane, including in the anti-center direction. Compared to \figref{fig:fullskymap}, the zodiacal light contamination is reduced by the continuum subtraction. However, some scan-aligned artifacts due to time-variable foregrounds remain, particularly toward the edges of the footprint where Galactic emission is fainter. In addition to zodiacal light, these may also arise from emission from the Earth's upper atmosphere~\citep{Hui_2026}.

To visualize the spatial relationship between PAH emission and ionized gas, \figref{fig:rgb} shows a two-color composite map of the Galactic plane, in which the  $3.3$-$\um$~PAH feature is assigned to the red channel and \bralpha\ emission to the blue channel. This rendering highlights the spatial complementarity of the two tracers, with PAH emission outlining extended shells surrounding compact regions of ionized gas. \figref{fig:rgb} also labels several prominent star-forming regions along the Galactic plane. On the left-hand side of the Galactic center, a broad bright area corresponds to the Cygnus complex, which shows strong emission in both tracers and was the focus of the \spherex\ wide-field study by \citet{Hora2026}. On the far left, well below the midplane, the Taurus region shows relatively strong PAH emission. Above the Galactic center, diffuse PAH emission is visible in the Ophiuchus region, while the \bralpha\ emission there is weaker and more localized. To the right of the Galactic center, the Vela region forms another bright enhancement in both tracers along the plane. Near the far-right side of the maps, slightly below the midplane, the Orion complex appears in both PAH and \bralpha\ emission. A zoomed-in view of the Orion Nebula is presented in \secref{ssec:orion}.

\subsection{A Zoomed View: The Orion Nebula}
\label{ssec:orion}

\begin{figure}
    \centering
    \includegraphics[width=1\linewidth]{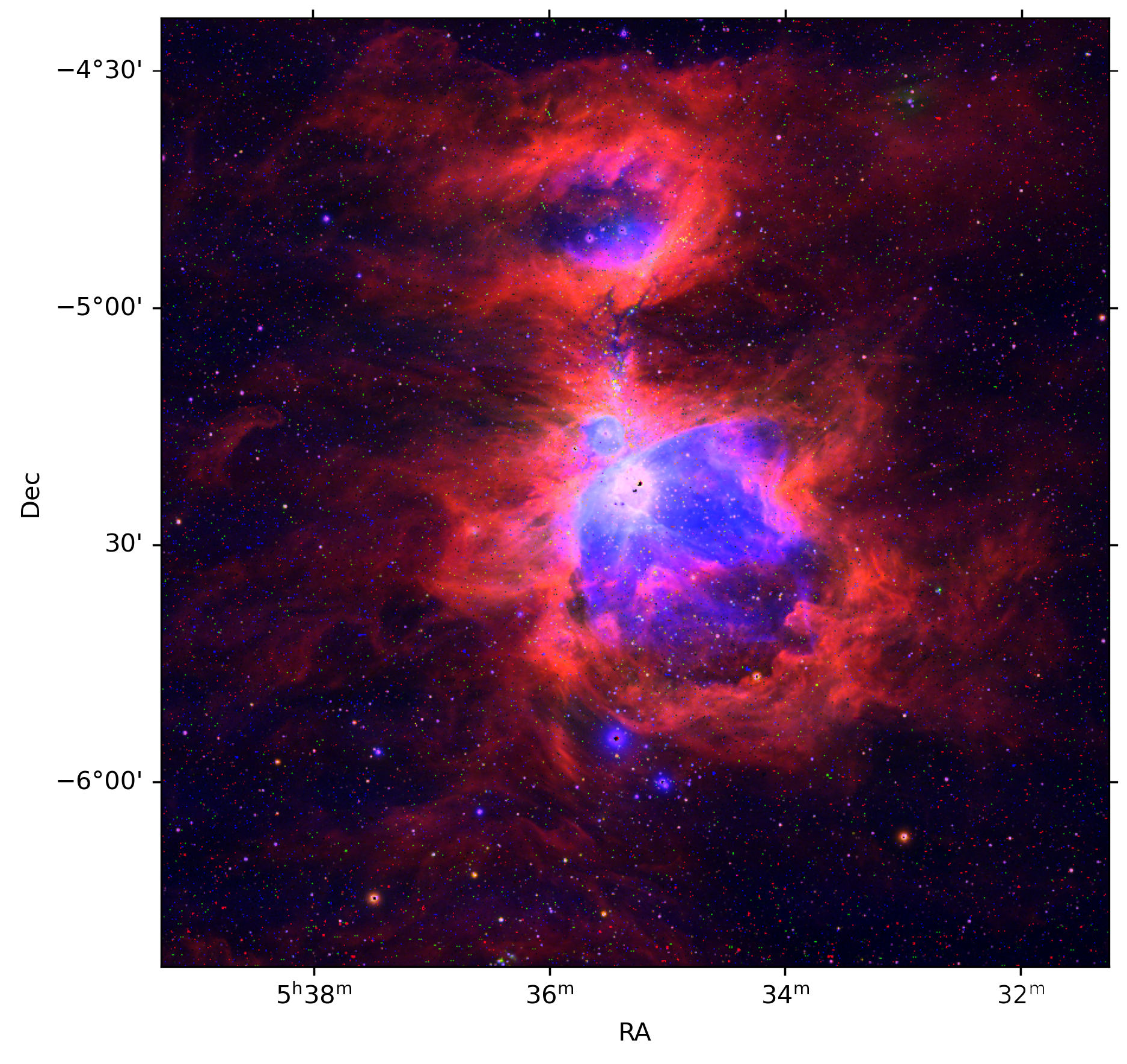}
    \caption{Composite RGB map of the Orion Nebula. Red shows $3.3$-$\um$~PAH emission, and blue shows $4.05$-$\um$~\bralpha, overlaid on a continuum background constructed from three channels spanning the \spherex\ wavelength range: 0.8~(blue), 2.5~(green), and $4.2~\um$~(red). The image illustrates the spatial separation between PAH emission and ionized gas at high resolution, consistent with the large-scale PAH depletion observed in the Galactic plane maps.}
    \label{fig:rgb_orion}
\end{figure}

To further illustrate the spatial relationship between PAH emission and ionized gas, \figref{fig:rgb_orion} presents a zoomed-in view of the Orion Nebula. The figure overlays the same two-color emission-line scheme as \figref{fig:rgb} on an RGB map of continuum emission constructed from three \spherex\ channels~(4.2, 2.5, and $0.8~\um$, respectively). This mosaic covers a $2^\circ \times 2^\circ$ field and is constructed by reprojecting the relevant \spherex\ exposures onto a Cartesian tangent-plane projection, rather than the \healpix\ representation, without applying the source masking and smoothing steps used for the diffuse-emission maps. The map has a pixel size of 9\arcsec, allowing us to resolve fine-scale structure within the nebula.

\figref{fig:rgb_orion} shows a spatial separation between PAHs~(red) and ionized gas~(blue) at the scale of an individual star-forming region, providing a high-resolution illustration of the processes studied at large scales across the Galactic plane. This is broadly consistent with the scenario in which PAHs are destroyed within the \HII\ region and excited in the surrounding PDR, producing the characteristic ring-like morphology. However, some PAH emission coincident with the ionized region is visible. This overlap arises from projection effects. The \HII\  region is an intrinsically three-dimensional object, so lines of sight through the projected interior of the nebula can still intercept PDR material on the near or far side of the bubble. More broadly, any line of sight through the Galactic plane integrates emission from multiple unrelated structures at different distances, so that PDR material can appear co-spatial with an \HII\ region in projection. Importantly, the spatial separation between PAH and ionized gas emission remains a robust large-scale phenomenon, clearly detectable despite this dilution, suggesting that PAH depletion in ionized environments is widespread across the Galactic plane.

\section{Map-space Correlations}
\label{sec:results}

In this section, we examine the relationships among PAH emission, ionized gas, and dust using pixel-by-pixel correlations. For consistency, all maps are smoothed to the same angular resolution with a $20\arcmin$ Gaussian kernel, rendered on a common pixel grid~($N_{\rm side}=512$), and restricted to the same sky area using consistent masks. Throughout this section, $I_\nu^{\rm PAH}$ denotes the emission at $3.3~\um$ and $I_\nu^{\rm Br\alpha}$ denotes the emission at $4.05~\um$.

Our primary goal is to understand how PAH abundance varies relative to the overall dust content and how it is affected by ionized environments. If PAH emission is powered by the same radiation field that heats the bulk of the dust, it should be correlated with the dust radiance. We therefore examine the correlation between the $3.3$-$\um$~PAH emission and the \planck\ dust radiance in \secref{ssec:corr857}. As described in \secref{sec:maps}, the diffuse maps show regions of reduced PAH emission in regions coincident with bright ionized-gas structures. To quantify this effect, we study correlations between PAH abundance and ionized gas traced by \bralpha\ in \secref{ssec:corrHII}.

\subsection{PAH Emission and Dust Radiance}
\label{ssec:corr857}

The dust radiance $\mathcal{R}$ measures the frequency-integrated dust emission and therefore traces both the dust column density and the strength of the radiation field heating the dust. Formally, it is defined as 
\begin{equation}
\mathcal{R} \equiv \int_0^\infty I_\nu^{\rm dust}\, d\nu,
\end{equation}
where~$I_\nu^{\rm dust}$~is the specific intensity of thermal dust emission. We employ the $\mathcal{R}$ map\footnote{COM\_CompMap\_Dust-GNILC-Radiance\_2048\_R2.00.fits, available from IRSA (\url{https://irsa.ipac.caltech.edu/data/Planck/release_2/all-sky-maps/previews/COM_CompMap_Dust-GNILC-Radiance_2048_R2.00/index.html})} of \citet{Planck2016}, who fit a modified blackbody model of dust emission in the Planck 353, 545, and 857\,GHz bands as well as the IRAS $100~\um$ band and then integrated this model over all frequencies. This map has an effective angular resolution of approximately $5\arcmin$. 

Because radiance measures the total power emitted by dust, it is proportional to the total power absorbed by dust. If a given line of sight has dust mass surface density $\Sigma_d$ and radiation field strength $U$, then $\mathcal{R} \propto U \Sigma_d$. PAH emission is produced predominantly by stochastically heated grains that cool down completely to their ground state before absorbing another photon~\citep{DraineLi2007, Richie2025}. Therefore, the total PAH emission at $3.3~\um$ is proportional to how often photons are absorbed by the grain and thus linearly proportional to $U$. If the mass fraction of dust in the form of PAHs is denoted $q_{\rm PAH}$, then $I_\nu^{\rm PAH} \propto U q_{\rm PAH} \Sigma_d$. Indeed, previous studies have shown that emission from small grains is tightly correlated with the dust radiance~\citep{Hensley2016}, suggesting that a similar relationship should be present for the $3.3$-$\um$~PAH emission mapped by SPHEREx.

\begin{figure} 
    \centering
    \includegraphics[width=1\linewidth]{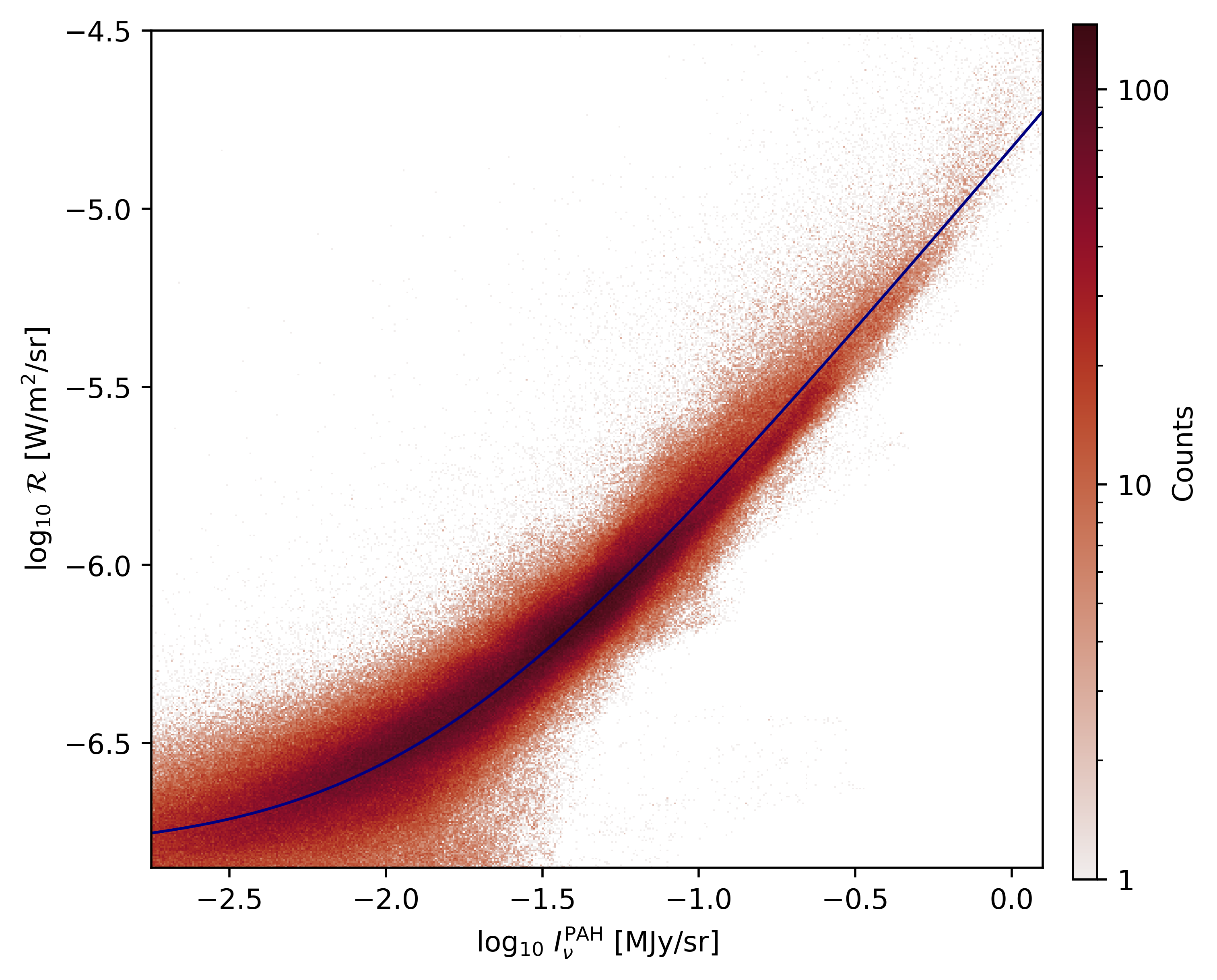}
    \caption{Pixel-by-pixel correlation between \spherex\ PAH intensity  $I_\nu^{\rm PAH}$ and \planck\ dust radiance $\mathcal{R}$.The line shows the linear model (Equation~\ref{eq:R_vs_PAH}). It guides the eye in order to accentuate deviations from a linear relationship. The overall correlation is strong, as expected for two tracers of dust emission, but systematic deviations from a simple power-law relation are visible at the highest PAH intensities, toward the upper right of the plot~($I_\nu^{\rm PAH} \gtrsim 0.56~\MJysr$), highlighting the distinct behavior of the $3.3$-$\um$~PAH emission on relatively dense lines of sight.}
    \label{fig:correlation}
\end{figure}

We correlate $I_\nu^{\rm PAH}$  with $\mathcal{R}$ on a pixel-by-pixel basis. As shown in \figref{fig:correlation}, we find a very strong positive correlation, with a Spearman coefficient of $\rho = 0.9340 \pm 0.0008$.

We use the Spearman rank correlation because it is less sensitive to outliers and bright sightlines and does not assume a linear relationship between the quantities. In \figref{fig:correlation}, the points show the binned distribution of pixel values, with color intensity indicating the density of points in each bin. The uncertainty is estimated using a permutation-based null test based on 100~random permutations of the Spearman ranks. This strong correlation serves as a powerful validation of our diffuse-emission map: it confirms that the $3.3$-$\um$ signal recovered after removing bright sources and neglecting zodiacal light contamination in the Galactic plane is dominated by dust-correlated emission on large scales, consistent with previous results showing that PAH abundance is strongly correlated with dust radiance~\citep{Hensley2016}.

Examining the relation shown in \figref{fig:correlation} in more detail, we find that it is close to linear over most of the dynamic range. We model the dust radiance as:
\begin{equation}
\mathcal{R} = A I_\nu^{\rm PAH} + B
\label{eq:R_vs_PAH}
\end{equation}
where $A$ and $B$ are fitting parameters. The model assumes a linear scaling between $\mathcal{R}$ and $I_\nu^{\rm PAH}$, consistent with the theoretical expectation. The nonzero offset $B$ absorbs contributions from the cosmic infrared background, zodiacal light, and interstellar light, and is expected since neither map has been corrected for all foreground and background components. The line is shown in \figref{fig:correlation} to guide the eye.

We note that at the highest PAH intensities~($I_\nu^{\rm PAH} \gtrsim 0.56~\MJysr$), the data systematically lie above the linear trend, indicating a deficit of $3.3$-$\um$~emission relative to that predicted from the total dust radiance. These deviations suggest that the $3.3$-$\um$~feature carries additional information about the local grain population or radiation environment beyond what is captured by the dust radiance alone. One likely reason is that the $3.3$-$\um$~feature is especially sensitive to the small, neutral PAH population~\citep{DraineLi2007}, while the dust radiance traces the total power absorbed and re-emitted by the broader dust population. Variations in PAH charge state, extinction of UV photons most critical for producing emission at $3.3~\um$, destruction of the smallest grains in hard radiation fields, and depletion of PAHs via coagulation onto larger grains in dense gas can all alter the $3.3$-$\um$ intensity relative to the total dust radiance. In this sense, the deviations from the linear trend likely reflect environmental changes in the PAH population and excitation conditions, including variations in $q_{\rm PAH}$ driven by local conditions such as radiation field hardness and metallicity.

\subsection{PAH Abundance and Ionized Hydrogen Emission}
\label{ssec:corrHII}

\begin{figure*}
    \centering
    \includegraphics[width=1\textwidth]{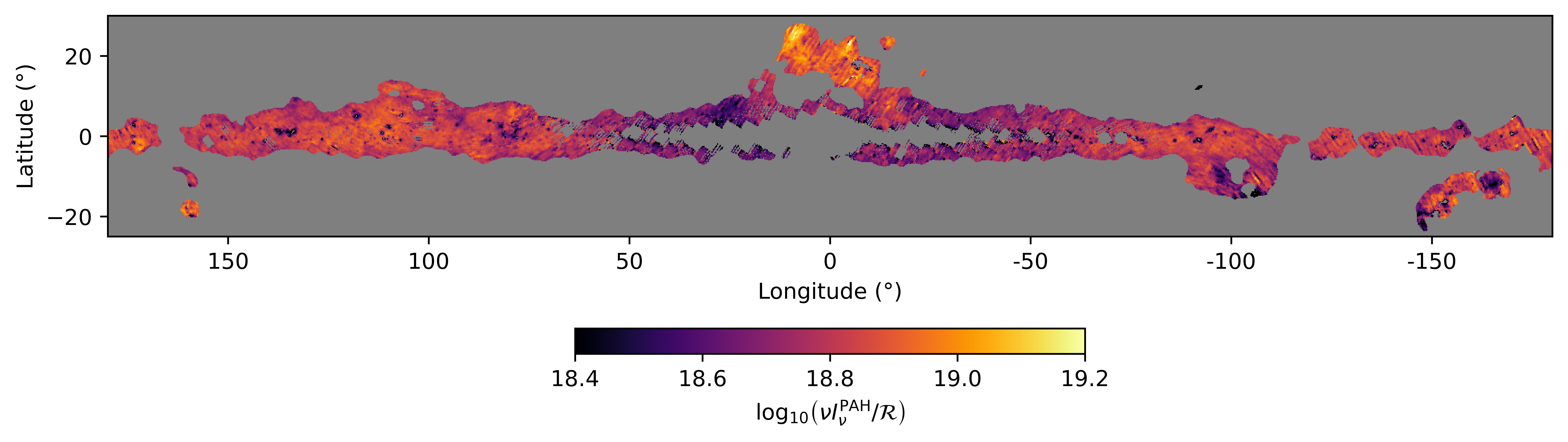}
    \caption{Radiance-normalized \spherex\ $3.3$-$\um$~PAH abundance map. In contrast to the PAH~emission map in \figref{fig:line_maps}, the PAH abundance highlights spatial variations relative to the dust column density.}
    \label{fig:pah_over_rad}
\end{figure*}

\begin{figure*}
    \centering
    \includegraphics[width=1\textwidth]{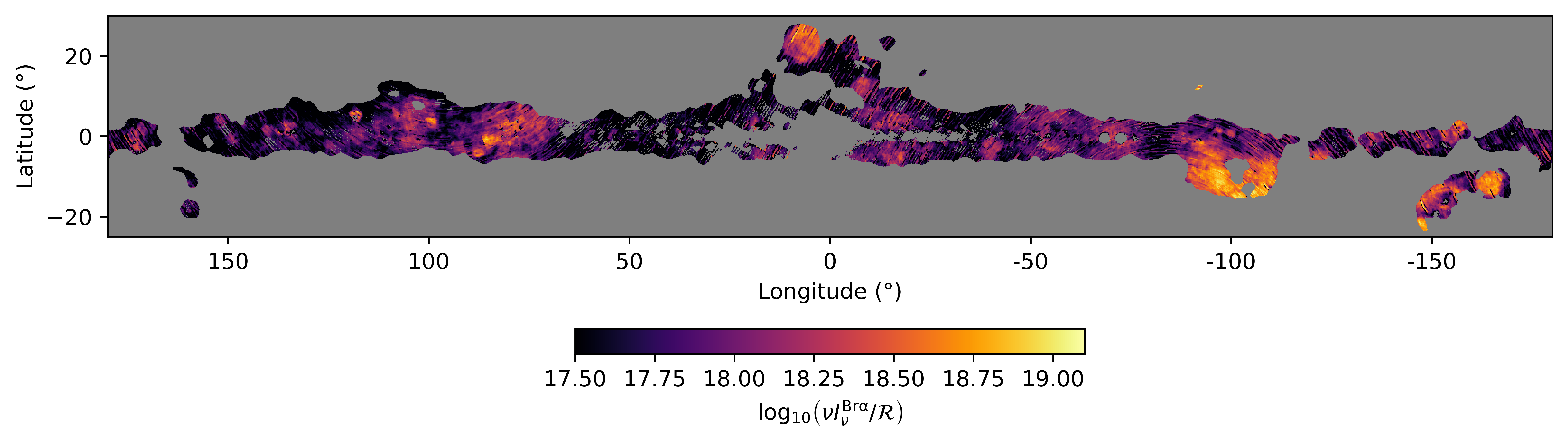}
    \caption{Radiance-normalized $4.05~\um$ \bralpha\ map, tracing ionized hydrogen emission relative to the interstellar radiation field.}
    \label{fig:h_over_rad}
\end{figure*}

To study the relationship between PAH emission and ionized gas, we compare the \spherex\ $3.3$-$\um$~PAH map with the \spherex\ \bralpha\ map, which traces diffuse ionized gas. To isolate intrinsic variations in PAH abundance, we normalize both quantities by the \planck\ dust radiance map, which accounts for differences in the local radiation field and total column density.
The radiance-normalized  PAH and \bralpha\ maps are shown in \figref{fig:pah_over_rad} and \figref{fig:h_over_rad}.

In the following, we first describe the radiance normalization and the resulting PAH abundance tracer~(\secref{ssec:corrHII_norm}), and then present evidence for PAH depletion from the pixel-by-pixel correlation~(\secref{ssec:corrHII_corr}). Extinction at these wavelengths is not expected to significantly affect the results, but we assess its potential impact in \secref{ssec:corrHII_ext}.

\subsubsection{Radiance Normalization and PAH Abundance}
\label{ssec:corrHII_norm}

As discussed in \secref{ssec:corr857}, dividing $I_\nu^{\rm PAH}$ by the radiance reduces its dependence on the radiation field strength and total column density. The radiance-normalized PAH intensity can therefore be interpreted as a tracer of the abundance of PAHs relative to the large dust grains responsible for the far-infrared emission across the Galactic plane. Since the $3.3$-$\um$~feature is primarily sensitive to the small, neutral PAHs, the radiance-normalized map traces the abundance of this population specifically, while the contribution of larger and more ionized PAHs is not directly constrained by this tracer alone.

Normalizing $I_\nu^{{\rm Br}\alpha}$ by the radiance serves as a proxy for the fraction of the gas column that is ionized. This normalization assumes a roughly constant gas-to-dust ratio, which may not hold across the entire Galactic plane due to metallicity variations. Therefore, variations in the global properties of the ISM could partially contribute to the spatial variations observed in the resulting map. Nevertheless, comparing these radiance-normalized intensities enables investigation of the effects of ionized gas on PAH abundance while mitigating correlations that are driven purely by the total column density or by the strength of the interstellar radiation field. 

To form radiance-normalized quantities, we multiply the \spherex\ specific intensity $I_\nu$ by the band frequency before dividing by the \planck\ dust radiance $\mathcal{R}$. Thus, we define the radiance-normalized maps as $\nu I_\nu / \mathcal{R}$. This construction is analogous to the $f_{\rm PAH}$ parameter introduced by \citet{Hensley2016}, defined as the ratio of $12~\um$~PAH emission to dust radiance and shown to trace the fraction of dust mass in PAHs.

We construct a mask that selects pixels above fixed surface-brightness thresholds in both the PAH and \bralpha\ maps and apply it to the radiance-normalized maps. Specifically, we require $I_\nu^{\rm PAH} > 0.063~\MJysr$ and $I_\nu^{\rm Br\alpha} > 0.025~\MJysr$. These thresholds represent a compromise between sky coverage and minimizing contamination from scan-aligned artifacts, which become more prominent at lower surface-brightness levels. Applying this mask restricts the pixel-by-pixel correlation analysis described below to regions where both tracers are reliably measured. The effects of more permissive masks are discussed in \secref{ssec:corrHII_corr}. With different masking thresholds, we find results that are broadly consistent.

Figures \ref{fig:pah_over_rad} and \ref{fig:h_over_rad} show the radiance-normalized PAH and \bralpha\ maps after applying the mask described above. The radiance-normalized PAH map traces the relative abundance of small, neutral PAHs with respect to the large dust grains. By removing the dependence on the local radiation field and total column density, it enables a view of where the $3.3$-$\um$~PAHs are preserved or depleted in different interstellar environments. 

The two maps show clearly different spatial distributions. The inferred PAH abundance is relatively diffuse along the Galactic plane, while the \bralpha\ emission is concentrated in compact regions. In many of the brightest \bralpha\ regions, the PAH abundance signal is comparatively weaker.

\subsubsection{PAH Depletion}
\label{ssec:corrHII_corr}

\begin{figure} 
    \centering
    \includegraphics[width=1\linewidth]{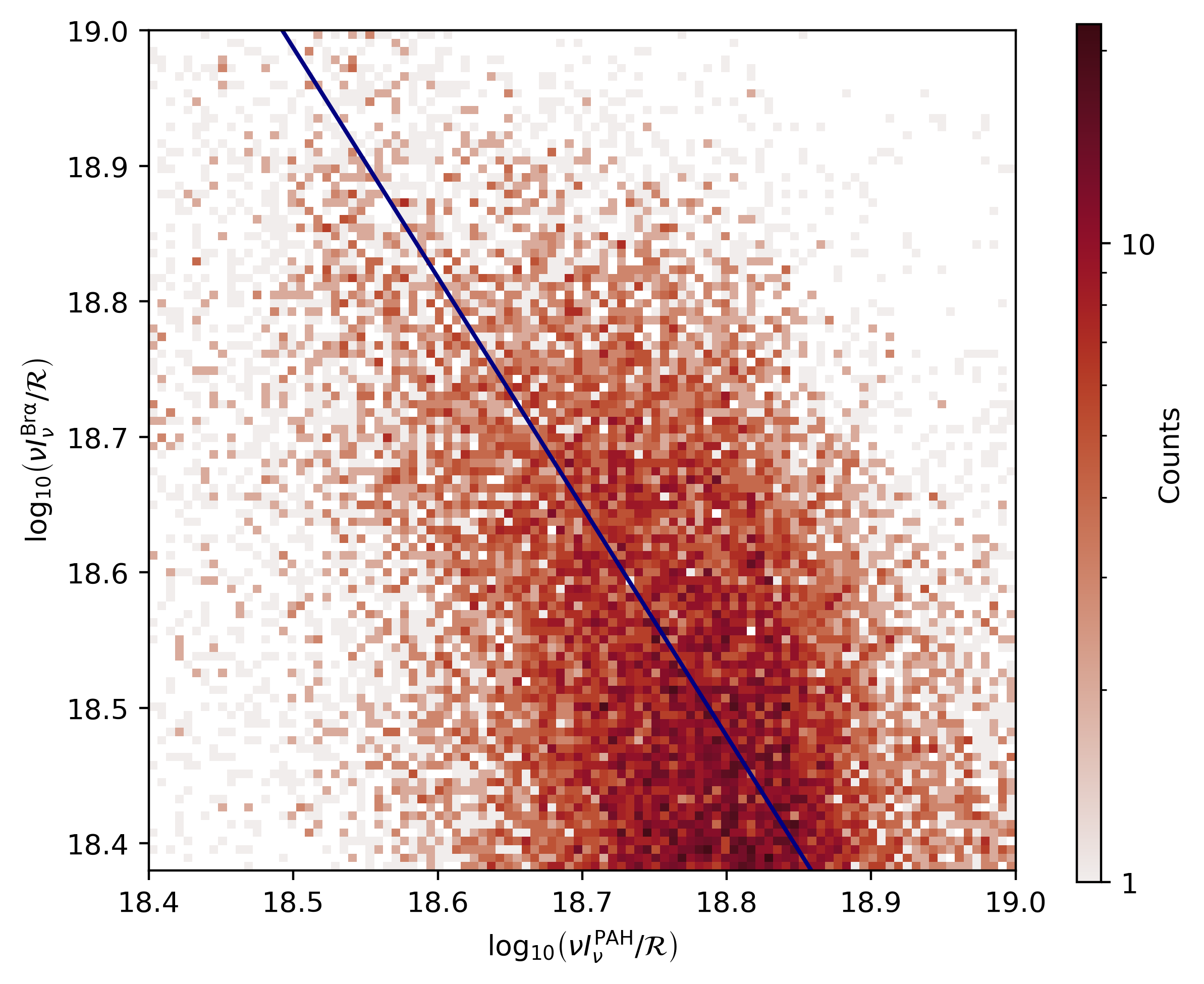}
    \caption{Pixel-scale correlation between radiance-normalized $3.3$-$\um$~PAH abundance and radiance-normalized \bralpha\ emission. The solid line shows an orthogonal distance regression fit with slope $m = -1.69$. The negative trend highlights the depletion of PAHs in regions of strong ionized hydrogen emission.}
    \label{fig:anticorrelation}
\end{figure}

\figref{fig:anticorrelation} shows the pixel-scale correlation between radiance-normalized PAH emission and radiance-normalized \bralpha\ emission. We calculate a Spearman correlation as described in \secref{ssec:corr857}, obtaining $\rho = -0.3661 \pm 0.0065$, indicating an anticorrelation. This result indicates a highly significant relationship between PAH abundance and radiance-normalized \bralpha\ emission. Despite the aggressive $20\arcmin$ smoothing, which suppresses small-scale structure and may dilute localized features such as individual \HII~regions and PDRs, the anticorrelation remains clearly detectable, indicating that PAH depletion occurs on large scales.

The substantial scatter in \figref{fig:anticorrelation} reflects a combination of effects. A likely dominant contribution is the line-of-sight superposition of multiple structures within the Galactic plane. Many PAH shells and ionized regions have complex three-dimensional geometries, and the observed emission at a given pixel can include contributions from several regions at different distances. Projection effects can therefore blur the transition between ionized gas and surrounding PDRs. Furthermore, the physical scale corresponding to the $20\arcmin$ smoothing varies significantly across the Galactic plane: in nearby regions such as the Orion Nebula, this corresponds to around 2~pc, while for more distant regions it can reach tens of parsecs, where the anticorrelation between PAH abundance and ionized gas may be diluted by averaging over larger physical scales. These projection and distance effects are therefore likely a primary driver of the scatter.

Scan-aligned artifacts arising from time-variable foregrounds may also contribute to the dispersion. Zodiacal-light emission is expected to be largely uncorrelated between channels, especially for those in different arrays, because the same sky regions are observed at different times. Although observations obtained within a few days could introduce a weak correlation, this would act against the observed anticorrelation. For an external tracer such as the radiance map, the zodiacal contribution is uncorrelated. To assess the impact of these artifacts, we repeat the analysis using more relaxed surface-brightness thresholds in the mask, which include sky areas where the artifacts are more prominent. We find that they increase the spread of data points but do not significantly alter the slope or strength of the anticorrelation. The primary effect of the artifacts is therefore to broaden the distribution and weaken the correlation rather than to produce spurious correlations between PAH and \bralpha\ emission. 

Overall, the anticorrelation indicates that regions of enhanced ionized hydrogen emission are, on average, associated with lower PAH abundance. Indeed, the strength of the correlation suggests that ionized gas is one of the dominant drivers of PAH abundance variations across the Galactic plane. This behavior is consistent with previously reported PAH depletion in regions of strong ionization~\citep[e.g.,][]{Churchwell2007,Paradis2011,Egorov2023,Egorov2025}. Our results extend these studies by demonstrating this anticorrelation across most of the Galactic plane using \spherex\ measurements of the 3.3~$\um$~PAH feature and \bralpha. 

However, reduced $3.3$-$\um$ emission does not necessarily imply PAH destruction. Part of the observed suppression may reflect changes in PAH charge state in ionized environments, where the hardness and intensity of the local radiation field depend on the spectral type of the exciting stars. In regions powered by later O-type or early B-type stars, for example, a weaker $3.3$-$\um$~feature could arise in part from a shift toward more ionized PAHs, even if some PAH material remains present. In addition, not all PAHs may be destroyed: the small PAHs responsible for the $3.3$-$\um$ emission are expected to be more vulnerable to destruction in harsher UV fields than larger PAHs. The observed trend likely reflects a combination of these effects.

\subsubsection{Dust Extinction}
\label{ssec:corrHII_ext}

We estimate the impact of dust extinction using the corrected Schlegel-Finkbeiner-Davis~(SFD) Galactic reddening map provided by \cite{Chiang2023}, which gives full-sky $E(B-V)$ values. This version removes a cosmic infrared background component that contaminates the original SFD map~\citep{Schlegel1998}, providing a slightly more accurate estimate of the Galactic reddening. Within our unmasked regions~(\figref{fig:pah_over_rad}), the median reddening is $E(B-V)=0.99$. For the Milky Way, we adopt the standard value $R_V=3.1$~\citep{Fitzpatrick2019}, and convert to visual extinction $A_V = R_V E(B-V)$. Using the extinction relation of \cite{Gordon2023}, which builds on the results of \cite{Gordon2009}, \cite{Gordon2021}, and \cite{Decleir2022}, we estimate the near-IR attenuation at our wavelengths, obtaining $A(3.3~\mu{\rm m}) \approx 0.15$~mag and $A(4.05~\mu{\rm m}) \approx 0.11$~mag. Therefore, we expect the extinction to have minimal impact on the measured correlations. To verify this directly, we repeat the correlation analysis after masking regions of high extinction. Specifically, we exclude all pixels with $E(B-V) > 2$ in the SFD reddening map. This threshold removes the highest-extinction lines of sight while retaining the majority of the Galactic plane. Recomputing the pixel-by-pixel correlation with this additional mask yields nearly identical results: the strength and slope of the anticorrelation change by only a few percent, and the overall trend remains clearly visible. This indicates that high-extinction regions do not drive the conclusions presented in \secref{ssec:corrHII_corr}.

\section{Discussion}
\label{sec:discussion}

These early studies demonstrate the unique capabilities of \spherex\ for mapping diffuse Galactic emission on large angular scales that are difficult to probe with targeted or narrow-field observations. \spherex's all-sky spectral coverage and sensitivity to near-infrared features allow us to trace PAHs and ionized gas across the Galaxy. The resulting maps provide a valuable new dataset for studying interstellar dust, radiation environments, and large-scale ISM structure, as illustrated by the results on PAH emission and dust correlation and PAH depletion in ionized regions presented in this paper.

Future work will build on these initial results in several important ways. Filtering zodiacal-light contamination will allow us to recover fainter diffuse emission at higher Galactic latitudes and eventually allow the construction of a full-sky PAH abundance map. As astrometric solutions improve for exposures in very dense stellar fields, we will be able to map and study PAH depletion in the complex environment of the Galactic center. Another future improvement will be to model the full spectral profile of the $3.3$-$\um$~PAH feature, rather than relying on the band excess above the local continuum, to obtain higher-fidelity measurements of PAH emission and abundance. We anticipate making the diffuse-emission maps publicly available after incorporating these improvements.

The \spherex\ diffuse-emission maps also enable a wide range of complementary studies. Comparison with longer-wavelength PAH tracers will determine how depletion strength depends on PAH size, while combining the $3.3$-$\um$ measurements with longer-wavelength dust tracers will provide a powerful template for constructing maps of Galactic dust extinction~\citep{Lee2025}. Measuring the variation of PAH abundance with galactocentric radius will separate local PAH processing near ionized regions from large-scale trends associated with the Galactic metallicity gradient~\citep{Whitcomb2024}. Mapping the $3.4~\um$ aliphatic feature will enable measurements of the aliphatic to aromatic ratio~\citep{Yang2016}. Additional analyses will explore correlations between PAH abundance and other tracers of the ISM, including anomalous microwave emission~\citep{Planck2011,Hensley2022,Pare2026}, to better understand the role of small dust grains in the ISM. Together, these developments demonstrate the potential of \spherex\ for future studies of diffuse emission and large-scale interstellar structure across the Galaxy.

\section{Conclusion}
\label{sec:conclusion}

In this work, we have presented large-scale maps of diffuse PAHs and ionized hydrogen emission constructed from early \spherex\ observations. The $3.3$-$\um$~map demonstrates that PAH emission is bright and detectable throughout the disk, and shows PAH-bright shells surrounding diffuse ionized gas, which trace PDRs over large angular scales.

We find a strong positive correlation between the $3.3$-$\um$~PAH intensity and the \planck\ dust radiance, confirming that the signal recovered is dominated by PAH emission on large scales. We construct a radiance-normalized PAH abundance map that isolates variations in the abundance of small, neutral PAHs from changes in the local radiation field. Using this data product, we identify an anticorrelation between $3.3$-$\um$~PAH abundance and ionized hydrogen traced by \bralpha, indicating systematic PAH depletion within ionized regions.

These findings extend previous studies of individual star-forming regions to Galactic scales and are consistent with extragalactic results, demonstrating that ionizing radiation is a dominant driver of $3.3$-$\um$~PAH abundance variations throughout the Milky Way.

\begin{acknowledgements}

We acknowledge support from the \spherex\ project under a contract from the NASA/Goddard Space Flight Center to the California Institute of Technology.

Part of the research described in this paper was carried out at the Jet Propulsion Laboratory, California Institute of Technology, under a contract with the National Aeronautics and Space Administration~(80NM0018D0004).

Part of this work was conducted during the Summer Undergraduate Research Fellowship~(SURF) program at the California Institute of Technology. We acknowledge support from the Samuel P. and Frances Krown SURF Fellowship during the summer of 2024 and the David L. Goodstein SURF Fellowship during the summer of 2025.

We also acknowledge the Texas Advanced Computing Center~(TACC) at The University of Texas at Austin for providing computational resources that have contributed to the research results reported within this paper.

Finally, we acknowledge the \spherex\ science and engineering teams for their contributions to the mission and development of the data products used in this work.

\end{acknowledgements}

\bibliography{main}{}

@ARTICLE{Bock2026,
       author = {{Bock}, James J. and {Aboobaker}, Asad M. and {Adamo}, Joseph and {Akeson}, Rachel and {Alred}, John M. and {Alibay}, Farah and {Ashby}, Matthew L.~N. and {Bach}, Yoonsoo P. and {Bleem}, Lindsey E. and {Bolton}, Douglas and {Braun}, David F. and {Bruton}, Sean and {Bryan}, Sean A. and {Chang}, Tzu-Ching and {Chen}, Shuang-Shuang and {Cheng}, Yun-Ting and {Cheshire}, IV, James R. and {Chiang}, Yi-Kuan and {de Janvry}, Jean Choppin and {Condon}, Samuel and {Cook}, Walter R. and {Cooray}, Asantha and {Crill}, Brendan P. and {Cukierman}, Ari J. and {Dor{\'e}}, Olivier and {Dowell}, C. Darren and {Dubois-Felsmann}, Gregory P. and {Eifler}, Tim and {Everett}, Spencer and {Fabinsky}, Beth E. and {Faisst}, Andreas L. and {Fanson}, James L. and {Farrington}, Allen H. and {Fatahi}, Tamim and {Fazar}, Candice M. and {Feder}, Richard M. and {Frater}, Eric H. and {Grasshorn Gebhardt}, Henry S. and {Giri}, Utkarsh and {Goldina}, Tatiana and {Gorjian}, Varoujan and {Habib}, Salman and {Hart}, William G. and {Heinrich}, Chen and {Hora}, Joseph L. and {Huai}, Zhaoyu and {Hui}, Howard and {Jo}, Young-Soo and {Jeong}, Woong-Seob and {Kang}, Jae Hwan and {Kang}, Miju and {Kecman}, Branislav and {Kim}, Chul-Hwan and {Kim}, Jaeyeong and {Kim}, Minjin and {Kim}, Young-Jun and {Kim}, Yongjung and {Kirkpatrick}, J. Davy and {Kobayashi}, Yosuke and {Korngut}, Phil M. and {Krause}, Elisabeth and {Lee}, Bomee and {Lee}, Ho-Gyu and {Lee}, Jae-Joon and {Lee}, Jeong-Eun and {Lisse}, Carey M. and {Mariani}, Giacomo and {Masters}, Daniel C. and {Mauskopf}, Philip D. and {Melnick}, Gary J. and {Minasyan}, Mary H. and {Mirocha}, Jordan and {Miyasaka}, Hiromasa and {Moore}, Anne and {Moore}, Bradley D. and {Murgia}, Giulia and {Naylor}, Bret J. and {Nelson}, Christina and {Nguyen}, Chi H. and {Nguyen}, Hien T. and {Noh}, Jinyoung K. and {Padin}, Stephen and {Paladini}, Roberta and {Park}, Sung-Joon and {Penanen}, Konstantin I. and {Putnam}, Dustin S. and {Pyo}, Jeonghyun and {Ramachandra}, Nesar and {Ramanathan}, Keshav and {Rustamkulov}, Zafar and {Reiley}, Daniel J. and {Rice}, Eric B. and {Rocca}, Jennifer M. and {Seok}, Ji Yeon and {Smith}, Roger and {Stober}, Jeremy and {Susca}, Sara and {Teplitz}, Harry I. and {Thelen}, Michael P. and {Tolls}, Volker and {Torrini}, Gabriela and {Trangsrud}, Amy R. and {Unwin}, Stephen and {Velicheti}, Phani and {Wang}, Pao-Yu and {Wen}, Robin Y. and {Werner}, Michael W. and {Williams}, Abby E. and {Williamson}, Ross and {Wincentsen}, James and {Windhorst}, Rogier A. and {Yang}, Soung-Chul and {Yang}, Yujin and {Zemcov}, Michael},
        title = "{The SPHEREx Satellite Mission}",
      journal = {\apj},
         year = 2026,
        month = mar,
       volume = {999},
       number = {1},
          eid = {139},
        pages = {139},
          doi = {10.3847/1538-4357/ae2be2},
archivePrefix = {arXiv},
       eprint = {2511.02985},
 primaryClass = {astro-ph.IM},
       adsurl = {https://ui.adsabs.harvard.edu/abs/2026ApJ...999..139B}
}

@ARTICLE{Crill2025,
       author = {{Crill}, Brendan P. and {Bach}, Yoonsoo P. and {Bryan}, Sean A. and {Choppin de Janvry}, Jean and {Cukierman}, Ari J. and {Dowell}, C. Darren and {Everett}, Spencer W. and {Fazar}, Candice and {Goldina}, Tatiana and {Huai}, Zhaoyu and {Hui}, Howard and {Jeong}, Woong-Seob and {Kang}, Jae Hwan and {Korngut}, Phillip M. and {Lee}, Jae Joon and {Masters}, Daniel C. and {Nguyen}, Chi H. and {Pyo}, Jeonghyun and {Symons}, Teresa and {Tolls}, Volker and {Yang}, Yujin and {Zemcov}, Michael and {Akeson}, Rachel and {Ashby}, Matthew L.~N. and {Bock}, James J. and {Chang}, Tzu-Ching and {Cheng}, Yun-Ting and {Chiang}, Yi-Kuan and {Cooray}, Asantha and {Dor{\'e}}, Olivier and {Faisst}, Andreas L. and {Feder}, Richard M. and {Werner}, Michael W.},
        title = "{The SPHEREx Sky Simulator: Science Data Modeling for the First All-sky Near-infrared Spectral Survey}",
      journal = {\apjs},
         year = 2025,
        month = nov,
       volume = {281},
       number = {1},
          eid = {10},
        pages = {10},
          doi = {10.3847/1538-4365/ae04cc},
archivePrefix = {arXiv},
       eprint = {2505.24856},
 primaryClass = {astro-ph.IM},
       adsurl = {https://ui.adsabs.harvard.edu/abs/2025ApJS..281...10C}
}

@ARTICLE{Bryan2025,
       author = {{Bryan}, Sean and {Bock}, James and {Burk}, Thomas and {Chang}, Tzu-Ching and {Crill}, Brendan P. and {Cukierman}, Ari and {Dore}, Olivier and {Dowell}, C. Darren and {Dubois-Felsmann}, Gregory and {Fabinsky}, Beth and {Hildebrandt-Rafels}, Sergi and {Hui}, Howard and {Hughes}, Kyle and {Korngut}, Phillip and {Mauskopf}, Philip and {Mena}, Julian and {Nguyen}, Chi and {Pourrahmani}, Milad and {Putnam}, Dustin and {Ramanathan}, Keshav and {Ridenhour}, Flora and {Roberson}, Cody and {Trangsrud}, Amy and {Unwin}, Stephen and {Wang}, Pao-Yu and {the SPHEREx Team}},
        title = "{Optimized Observation Sequencing in Low-Earth Orbit with the SPHEREx Survey Planning Software}",
      journal = {arXiv e-prints},
         year = 2025,
        month = aug,
          eid = {arXiv:2508.20332},
        pages = {arXiv:2508.20332},
          doi = {10.48550/arXiv.2508.20332},
archivePrefix = {arXiv},
       eprint = {2508.20332},
 primaryClass = {astro-ph.IM},
       adsurl = {https://ui.adsabs.harvard.edu/abs/2025arXiv250820332B}
}

@ARTICLE{Akeson2025,
       author = {{Akeson}, Rachel and {Dubois-Felsmann}, Gregory P. and {Crill}, Brendan P. and {Faisst}, Andreas L. and {Fatahi}, Tamim and {Fazar}, Candice M. and {Goldina}, Tatiana and {Masters}, Daniel C. and {Nelson}, Christina and {Paladini}, Roberta and {Teplitz}, Harry I. and {Torrini}, Gabriela and {Velicheti}, Phani and {Ashby}, Matthew L.~N. and {Avner}, Dan and {Bach}, Yoonsoo P. and {Bock}, James J. and {Bruton}, Sean and {Bryan}, Sean A. and {Chang}, Tzu-Ching and {Chen}, Shuang-Shuang and {Cukierman}, Ari J. and {Dore}, O. and {Dowell}, C. Darren and {Everett}, Spencer and {Feder}, Richard M. and {Huai}, Zhaoyu and {Hui}, Howard and {Jeong}, Woong-Seob and {Jo}, Young-Soo and {Korngut}, Phil M. and {Kwon}, Yuna G. and {Lee}, Bomee and {Melnick}, Gary J. and {Murgia}, Giulia and {Nguyen}, Chi H. and {Pourrahmani}, Milad and {Rustamkulov}, Zafar and {Tolls}, Volker and {Wang}, Pao-Yu and {Yang}, Yujin and {Zemcov}, Michael},
        title = "{The SPHEREx Image and Spectrophotometry Processing Pipeline}",
      journal = {arXiv e-prints},
         year = 2025,
        month = nov,
          eid = {arXiv:2511.15823},
        pages = {arXiv:2511.15823},
          doi = {10.48550/arXiv.2511.15823},
archivePrefix = {arXiv},
       eprint = {2511.15823},
 primaryClass = {astro-ph.IM},
       adsurl = {https://ui.adsabs.harvard.edu/abs/2025arXiv251115823A}
}

@ARTICLE{Hui2026,
       author = {{Hui}, Howard and {Bock}, James J. and {Condon}, Samuel and
                 {Dowell}, C. Darren and {Jeong}, Woong-Seob and {Jo}, Young-soo and
                 {Korngut}, Phil M. and {Manatt}, Kenneth and {Nguyen}, Chi and
                 {Nguyen}, Hien and {Padin}, Stephen and {Park}, Sung-Joon and
                 {Pyo}, Jeonghyun and {Yang}, Yujin and {Ashby}, Matthew L.~N. and
                 {Bach}, Yoonsoo P. and {Chang}, Tzu-Ching and {Cheng}, Yun-Ting and
                 {Chiang}, Yi-Kuan and {Cooray}, Asantha and {Crill}, Brendan P. and
                 {Cukierman}, Ari J. and {Dor{\'e}}, Olivier and {Faisst}, Andreas L. and
                 {Hora}, Joseph L. and {Kang}, Jae Hwan and {Lee}, BoMee and
                 {Lisse}, Carey M. and {Masters}, Daniel C. and {Paladini}, Roberta and
                 {Rustamkulov}, Zafar and {Tolls}, Volker and {Werner}, Michael W. and
                 {Zemcov}, Michael},
        title = "{Spectral Response of SPHEREx}",
      journal = {\apjs},
         year = 2026,
        month = apr,
       volume = {284},
       number = {1},
          eid = {10},
        pages = {10},
          doi = {10.3847/1538-4365/ae522c},
archivePrefix = {arXiv},
       eprint = {2602.09139},
 primaryClass = {astro-ph.IM},
       adsurl = {https://ui.adsabs.harvard.edu/abs/2026ApJS..284...10H}
}

@ARTICLE{Hora2026,
       author = {{Hora}, Joseph L. and {Noh}, Jinyoung K. and {Melnick}, Gary J. and {Hensley}, Brandon S. and {Paladini}, Roberta and {Lee}, Jeong-Eun and {Ashby}, Matthew L.~N. and {Tolls}, Volker and {Kim}, Jaeyeong and {Werner}, Michael W. and {Bock}, James J. and {Bruton}, Sean and {Chen}, Shuang-Shuang and {Chang}, Tzu-Ching and {Chiang}, Yi-Kuan and {Cooray}, Asantha and {Crill}, Brendan P. and {Cukierman}, Ari J. and {Dor{\'e}}, Olivier and {Faisst}, Andreas L. and {Huai}, Zhaoyu and {Hui}, Howard and {Jeong}, Woong-Seob and {Kang}, Miju and {Korngut}, Phil M. and {Lee}, Ho-Gyu and {Lisse}, Carey M. and {Masters}, Daniel C. and {Murgia}, Giulia and {Nguyen}, Chi H. and {Rustamkulov}, Zafar and {Seok}, Ji Yeon and {Wen}, Robin Y. and {Yang}, Yujin and {Zemcov}, Michael},
        title = "{SPHEREx Widefield Infrared Spectral Mapping of Interstellar Ices and Polycyclic Aromatic Hydrocarbons}",
      journal = {\apj},
         year = 2026,
        month = apr,
       volume = {1001},
       number = {2},
          eid = {165},
        pages = {165},
          doi = {10.3847/1538-4357/ae5180},
archivePrefix = {arXiv},
       eprint = {2603.12390},
 primaryClass = {astro-ph.GA},
       adsurl = {https://ui.adsabs.harvard.edu/abs/2026ApJ..1001..165H}
}

@ARTICLE{Wright2010,
       author = {{Wright}, Edward L. and {Eisenhardt}, Peter R.~M. and {Mainzer}, Amy K. and {Ressler}, Michael E. and {Cutri}, Roc M. and {Jarrett}, Thomas and {Kirkpatrick}, J. Davy and {Padgett}, Deborah and {McMillan}, Robert S. and {Skrutskie}, Michael and {Stanford}, S.~A. and {Cohen}, Martin and {Walker}, Russell G. and {Mather}, John C. and {Leisawitz}, David and {Gautier}, III, Thomas N. and {McLean}, Ian and {Benford}, Dominic and {Lonsdale}, Carol J. and {Blain}, Andrew and {Mendez}, Bryan and {Irace}, William R. and {Duval}, Valerie and {Liu}, Fengchuan and {Royer}, Don and {Heinrichsen}, Ingolf and {Howard}, Joan and {Shannon}, Mark and {Kendall}, Martha and {Walsh}, Amy L. and {Larsen}, Mark and {Cardon}, Joel G. and {Schick}, Scott and {Schwalm}, Mark and {Abid}, Mohamed and {Fabinsky}, Beth and {Naes}, Larry and {Tsai}, Chao-Wei},
        title = "{The Wide-field Infrared Survey Explorer (WISE): Mission Description and Initial On-orbit Performance}",
      journal = {\aj},
         year = 2010,
        month = dec,
       volume = {140},
       number = {6},
        pages = {1868-1881},
          doi = {10.1088/0004-6256/140/6/1868},
archivePrefix = {arXiv},
       eprint = {1008.0031},
 primaryClass = {astro-ph.IM},
       adsurl = {https://ui.adsabs.harvard.edu/abs/2010AJ....140.1868W}
}

@ARTICLE{Werner2004,
       author = {{Werner}, M.~W. and {Roellig}, T.~L. and {Low}, F.~J. and {Rieke}, G.~H. and {Rieke}, M. and {Hoffmann}, W.~F. and {Young}, E. and {Houck}, J.~R. and {Brandl}, B. and {Fazio}, G.~G. and {Hora}, J.~L. and {Gehrz}, R.~D. and {Helou}, G. and {Soifer}, B.~T. and {Stauffer}, J. and {Keene}, J. and {Eisenhardt}, P. and {Gallagher}, D. and {Gautier}, T.~N. and {Irace}, W. and {Lawrence}, C.~R. and {Simmons}, L. and {Van Cleve}, J.~E. and {Jura}, M. and {Wright}, E.~L. and {Cruikshank}, D.~P.},
        title = "{The Spitzer Space Telescope Mission}",
      journal = {\apjs},
         year = 2004,
        month = sep,
       volume = {154},
       number = {1},
        pages = {1-9},
          doi = {10.1086/422992},
archivePrefix = {arXiv},
       eprint = {astro-ph/0406223},
 primaryClass = {astro-ph},
       adsurl = {https://ui.adsabs.harvard.edu/abs/2004ApJS..154....1W}
}

@ARTICLE{Sandstrom2010,
       author = {{Sandstrom}, Karin M. and {Bolatto}, Alberto D. and {Draine}, B.~T. and {Bot}, Caroline and {Stanimirovi{\'c}}, Sne{\v{z}}ana},
        title = "{The Spitzer Survey of the Small Magellanic Cloud (S$^{3}$MC): Insights into the Life Cycle of Polycyclic Aromatic Hydrocarbons}",
      journal = {\apj},
         year = 2010,
        month = jun,
       volume = {715},
       number = {2},
        pages = {701-723},
          doi = {10.1088/0004-637X/715/2/701},
archivePrefix = {arXiv},
       eprint = {1003.4516},
 primaryClass = {astro-ph.CO},
       adsurl = {https://ui.adsabs.harvard.edu/abs/2010ApJ...715..701S}
}

@ARTICLE{Churchwell2007,
       author = {{Churchwell}, E. and {Watson}, D.~F. and {Povich}, M.~S. and {Taylor}, M.~G. and {Babler}, B.~L. and {Meade}, M.~R. and {Benjamin}, R.~A. and {Indebetouw}, R. and {Whitney}, B.~A.},
        title = "{The Bubbling Galactic Disk. II. The Inner 20{\textdegree}}",
      journal = {\apj},
         year = 2007,
        month = nov,
       volume = {670},
       number = {1},
        pages = {428-441},
          doi = {10.1086/521646},
       adsurl = {https://ui.adsabs.harvard.edu/abs/2007ApJ...670..428C}
}

@ARTICLE{Paradis2011,
       author = {{Paradis}, D{\'e}borah and {Paladini}, Roberta and {Noriega-Crespo}, Alberto and {Lagache}, Guilaine and {Kawamura}, Akiko and {Onishi}, Toshikazu and {Fukui}, Yasuo},
        title = "{Spitzer Characterization of Dust in the Ionized Medium of the Large Magellanic Cloud}",
      journal = {\apj},
         year = 2011,
        month = jul,
       volume = {735},
       number = {1},
          eid = {6},
        pages = {6},
          doi = {10.1088/0004-637X/735/1/6},
archivePrefix = {arXiv},
       eprint = {1104.1098},
 primaryClass = {astro-ph.GA},
       adsurl = {https://ui.adsabs.harvard.edu/abs/2011ApJ...735....6P}
}

@ARTICLE{Watson2010,
       author = {{Watson}, C. and {Hanspal}, U. and {Mengistu}, A.},
        title = "{Triggered Star Formation and Dust Around Mid-infrared-identified Bubbles}",
      journal = {\apj},
         year = 2010,
        month = jun,
       volume = {716},
       number = {2},
        pages = {1478-1492},
          doi = {10.1088/0004-637X/716/2/1478},
archivePrefix = {arXiv},
       eprint = {1006.0206},
 primaryClass = {astro-ph.GA},
       adsurl = {https://ui.adsabs.harvard.edu/abs/2010ApJ...716.1478W}
}

@ARTICLE{Watson2009,
       author = {{Watson}, C. and {Corn}, T. and {Churchwell}, E.~B. and {Babler}, B.~L. and {Povich}, M.~S. and {Meade}, M.~R. and {Whitney}, B.~A.},
        title = "{IR Dust Bubbles. II. Probing the Detailed Structure and Young Massive Stellar Populations of Galactic H II Regions}",
      journal = {\apj},
         year = 2009,
        month = mar,
       volume = {694},
       number = {1},
        pages = {546-555},
          doi = {10.1088/0004-637X/694/1/546},
archivePrefix = {arXiv},
       eprint = {0901.1097},
 primaryClass = {astro-ph.GA},
       adsurl = {https://ui.adsabs.harvard.edu/abs/2009ApJ...694..546W}
}

@ARTICLE{Winston2012,
       author = {{Winston}, E. and {Wolk}, S.~J. and {Bourke}, T.~L. and {Megeath}, S.~T. and {Gutermuth}, R. and {Spitzbart}, B.},
        title = "{Spitzer Observations of Bow Shocks and Outflows in RCW 38}",
      journal = {\apj},
         year = 2012,
        month = jan,
       volume = {744},
       number = {2},
          eid = {126},
        pages = {126},
          doi = {10.1088/0004-637X/744/2/126},
archivePrefix = {arXiv},
       eprint = {1111.4413},
 primaryClass = {astro-ph.SR},
       adsurl = {https://ui.adsabs.harvard.edu/abs/2012ApJ...744..126W}
}

@ARTICLE{Allamandola1985,
       author = {{Allamandola}, L.~J. and {Tielens}, A.~G.~G.~M. and {Barker}, J.~R.},
        title = "{Polycyclic aromatic hydrocarbons and the unidentified infrared emission bands: auto exhaust along the milky way.}",
      journal = {\apjl},
         year = 1985,
        month = mar,
       volume = {290},
        pages = {L25-L28},
          doi = {10.1086/184435},
       adsurl = {https://ui.adsabs.harvard.edu/abs/1985ApJ...290L..25A}
}

@ARTICLE{Tielens2008,
       author = {{Tielens}, A.~G.~G.~M.},
        title = "{Interstellar polycyclic aromatic hydrocarbon molecules.}",
      journal = {\araa},
         year = 2008,
        month = sep,
       volume = {46},
        pages = {289-337},
          doi = {10.1146/annurev.astro.46.060407.145211},
       adsurl = {https://ui.adsabs.harvard.edu/abs/2008ARA&A..46..289T}
}

@ARTICLE{LegerPuget1984,
       author = {{Leger}, A. and {Puget}, J.~L.},
        title = "{Identification of the Unidentified Infrared Emission Features of Interstellar Dust}",
      journal = {\aap},
         year = 1984,
        month = aug,
       volume = {137},
        pages = {L5-L8},
       adsurl = {https://ui.adsabs.harvard.edu/abs/1984A&A...137L...5L}
}

@ARTICLE{Smith2007,
       author = {{Smith}, J.~D.~T. and {Draine}, B.~T. and {Dale}, D.~A. and {Moustakas}, J. and {Kennicutt}, Jr., R.~C. and {Helou}, G. and {Armus}, L. and {Roussel}, H. and {Sheth}, K. and {Bendo}, G.~J. and {Buckalew}, B.~A. and {Calzetti}, D. and {Engelbracht}, C.~W. and {Gordon}, K.~D. and {Hollenbach}, D.~J. and {Li}, A. and {Malhotra}, S. and {Murphy}, E.~J. and {Walter}, F.},
        title = "{The Mid-Infrared Spectrum of Star-forming Galaxies: Global Properties of Polycyclic Aromatic Hydrocarbon Emission}",
      journal = {\apj},
         year = 2007,
        month = feb,
       volume = {656},
       number = {2},
        pages = {770-791},
          doi = {10.1086/510549},
archivePrefix = {arXiv},
       eprint = {astro-ph/0610913},
 primaryClass = {astro-ph},
       adsurl = {https://ui.adsabs.harvard.edu/abs/2007ApJ...656..770S}
}

@ARTICLE{Galliano2008,
       author = {{Galliano}, Fr{\'e}d{\'e}ric and {Madden}, Suzanne C. and {Tielens}, Alexander G.~G.~M. and {Peeters}, Els and {Jones}, Anthony P.},
        title = "{Variations of the Mid-IR Aromatic Features inside and among Galaxies}",
      journal = {\apj},
         year = 2008,
        month = may,
       volume = {679},
       number = {1},
        pages = {310-345},
          doi = {10.1086/587051},
archivePrefix = {arXiv},
       eprint = {0801.4955},
 primaryClass = {astro-ph},
       adsurl = {https://ui.adsabs.harvard.edu/abs/2008ApJ...679..310G}
}

@ARTICLE{PereiraSantaella2010,
       author = {{Pereira-Santaella}, Miguel and {Alonso-Herrero}, Almudena and {Rieke}, George H. and {Colina}, Luis and {D{\'\i}az-Santos}, Tanio and {Smith}, J.-D.~T. and {P{\'e}rez-Gonz{\'a}lez}, Pablo G. and {Engelbracht}, Charles W.},
        title = "{Local Luminous Infrared Galaxies. I. Spatially Resolved Observations with the Spitzer Infrared Spectrograph}",
      journal = {\apjs},
         year = 2010,
        month = jun,
       volume = {188},
       number = {2},
        pages = {447-472},
          doi = {10.1088/0067-0049/188/2/447},
archivePrefix = {arXiv},
       eprint = {1004.1364},
 primaryClass = {astro-ph.GA},
       adsurl = {https://ui.adsabs.harvard.edu/abs/2010ApJS..188..447P}
}

@ARTICLE{Sloan1997,
       author = {{Sloan}, G.~C. and {Bregman}, J.~D. and {Geballe}, T.~R. and {Allamandola}, L.~J. and {Woodward}, E.},
        title = "{Variations in the 3 Micron Spectrum across the Orion Bar: Polycyclic Aromatic Hydrocarbons and Related Molecules}",
      journal = {\apj},
         year = 1997,
        month = jan,
       volume = {474},
       number = {2},
        pages = {735-740},
          doi = {10.1086/303484},
       adsurl = {https://ui.adsabs.harvard.edu/abs/1997ApJ...474..735S}
}

@ARTICLE{Mori2014,
       author = {{Mori}, Tamami I. and {Onaka}, Takashi and {Sakon}, Itsuki and {Ishihara}, Daisuke and {Shimonishi}, Takashi and {Ohsawa}, Ryou and {Bell}, Aaron C.},
        title = "{Observational Studies on the Near-infrared Unidentified Emission Bands in Galactic H II Regions}",
      journal = {\apj},
         year = 2014,
        month = mar,
       volume = {784},
       number = {1},
          eid = {53},
        pages = {53},
          doi = {10.1088/0004-637X/784/1/53},
archivePrefix = {arXiv},
       eprint = {1401.7879},
 primaryClass = {astro-ph.GA},
       adsurl = {https://ui.adsabs.harvard.edu/abs/2014ApJ...784...53M}
}

@ARTICLE{WeingartnerDraine2001,
       author = {{Weingartner}, Joseph C. and {Draine}, B.~T.},
        title = "{Photoelectric Emission from Interstellar Dust: Grain Charging and Gas Heating}",
      journal = {\apjs},
         year = 2001,
        month = jun,
       volume = {134},
       number = {2},
        pages = {263-281},
          doi = {10.1086/320852},
archivePrefix = {arXiv},
       eprint = {astro-ph/9907251},
 primaryClass = {astro-ph},
       adsurl = {https://ui.adsabs.harvard.edu/abs/2001ApJS..134..263W}
}

@ARTICLE{Rigopoulou2021,
       author = {{Rigopoulou}, D. and {Donnan}, F.~R. and {Garc{\'\i}a-Bernete}, I. and {Pereira-Santaella}, M. and {Alonso-Herrero}, A. and {Davies}, R. and {Hunt}, L.~K. and {Roche}, P.~F. and {Shimizu}, T.},
        title = "{Polycyclic aromatic hydrocarbon emission in galaxies as seen with JWST}",
      journal = {\mnras},
         year = 2024,
        month = aug,
       volume = {532},
       number = {2},
        pages = {1598-1611},
          doi = {10.1093/mnras/stae1535},
archivePrefix = {arXiv},
       eprint = {2406.11415},
 primaryClass = {astro-ph.GA},
       adsurl = {https://ui.adsabs.harvard.edu/abs/2024MNRAS.532.1598R}
}

@ARTICLE{DraineLi2007,
       author = {{Draine}, B.~T. and {Li}, Aigen},
        title = "{Infrared Emission from Interstellar Dust. IV. The Silicate-Graphite-PAH Model in the Post-Spitzer Era}",
      journal = {\apj},
         year = 2007,
        month = mar,
       volume = {657},
       number = {2},
        pages = {810-837},
          doi = {10.1086/511055},
archivePrefix = {arXiv},
       eprint = {astro-ph/0608003},
 primaryClass = {astro-ph},
       adsurl = {https://ui.adsabs.harvard.edu/abs/2007ApJ...657..810D}
}

@ARTICLE{Hensley2016,
       author = {{Hensley}, Brandon S. and {Draine}, B.~T. and {Meisner}, Aaron M.},
        title = "{A Case Against Spinning PAHs as the Source of the Anomalous Microwave Emission}",
      journal = {\apj},
         year = 2016,
        month = aug,
       volume = {827},
       number = {1},
          eid = {45},
        pages = {45},
          doi = {10.3847/0004-637X/827/1/45},
archivePrefix = {arXiv},
       eprint = {1505.02157},
 primaryClass = {astro-ph.GA},
       adsurl = {https://ui.adsabs.harvard.edu/abs/2016ApJ...827...45H}
}

@ARTICLE{Garay1999,
       author = {{Garay}, Guido and {Lizano}, Susana},
        title = "{Massive Stars: Their Environment and Formation}",
      journal = {\pasp},
         year = 1999,
        month = sep,
       volume = {111},
       number = {763},
        pages = {1049-1087},
          doi = {10.1086/316416},
archivePrefix = {arXiv},
       eprint = {astro-ph/9907293},
 primaryClass = {astro-ph},
       adsurl = {https://ui.adsabs.harvard.edu/abs/1999PASP..111.1049G}
}

@ARTICLE{Moises2011,
       author = {{Mois{\'e}s}, A.~P. and {Damineli}, A. and {Figuer{\^e}do}, E. and {Blum}, R.~D. and {Conti}, P.~S. and {Barbosa}, C.~L.},
        title = "{Spectrophotometric distances to Galactic H II regions}",
      journal = {\mnras},
         year = 2011,
        month = feb,
       volume = {411},
       number = {2},
        pages = {705-760},
          doi = {10.1111/j.1365-2966.2010.17713.x},
archivePrefix = {arXiv},
       eprint = {1009.3924},
 primaryClass = {astro-ph.SR},
       adsurl = {https://ui.adsabs.harvard.edu/abs/2011MNRAS.411..705M}
}

@ARTICLE{Yano2016,
       author = {{Yano}, Kenichi and {Nakagawa}, Takao and {Isobe}, Naoki and {Shirahata}, Mai},
        title = "{Star Formation in Ultraluminous Infrared Galaxies Probed with AKARI Near-infrared Spectroscopy}",
      journal = {\apj},
         year = 2016,
        month = dec,
       volume = {833},
       number = {2},
          eid = {272},
        pages = {272},
          doi = {10.3847/1538-4357/833/2/272},
archivePrefix = {arXiv},
       eprint = {1611.00006},
 primaryClass = {astro-ph.GA},
       adsurl = {https://ui.adsabs.harvard.edu/abs/2016ApJ...833..272Y}
}

@ARTICLE{Chiang2023,
       author = {{Chiang}, Yi-Kuan},
        title = "{Corrected SFD: A More Accurate Galactic Dust Map with Minimal Extragalactic Contamination}",
      journal = {\apj},
         year = 2023,
        month = dec,
       volume = {958},
       number = {2},
          eid = {118},
        pages = {118},
          doi = {10.3847/1538-4357/acf4a1},
archivePrefix = {arXiv},
       eprint = {2306.03926},
 primaryClass = {astro-ph.GA},
       adsurl = {https://ui.adsabs.harvard.edu/abs/2023ApJ...958..118C}
}

@ARTICLE{Fitzpatrick2019,
       author = {{Fitzpatrick}, E.~L. and {Massa}, Derck and {Gordon}, Karl D. and {Bohlin}, Ralph and {Clayton}, Geoffrey C.},
        title = "{An Analysis of the Shapes of Interstellar Extinction Curves. VII. Milky Way Spectrophotometric Optical-through-ultraviolet Extinction and Its R-dependence}",
      journal = {\apj},
         year = 2019,
        month = dec,
       volume = {886},
       number = {2},
          eid = {108},
        pages = {108},
          doi = {10.3847/1538-4357/ab4c3a},
archivePrefix = {arXiv},
       eprint = {1910.08852},
 primaryClass = {astro-ph.GA},
       adsurl = {https://ui.adsabs.harvard.edu/abs/2019ApJ...886..108F}
}

@ARTICLE{Gordon2009,
       author = {{Gordon}, Karl D. and {Cartledge}, Stefan and {Clayton}, Geoffrey C.},
        title = "{FUSE Measurements of Far-Ultraviolet Extinction. III. The Dependence on R(V) and Discrete Feature Limits from 75 Galactic Sightlines}",
      journal = {\apj},
         year = 2009,
        month = nov,
       volume = {705},
       number = {2},
        pages = {1320-1335},
          doi = {10.1088/0004-637X/705/2/1320},
archivePrefix = {arXiv},
       eprint = {0909.3087},
 primaryClass = {astro-ph.GA},
       adsurl = {https://ui.adsabs.harvard.edu/abs/2009ApJ...705.1320G}
}

@ARTICLE{Gordon2021,
       author = {{Gordon}, Karl D. and {Misselt}, Karl A. and {Bouwman}, Jeroen and {Clayton}, Geoffrey C. and {Decleir}, Marjorie and {Hines}, Dean C. and {Pendleton}, Yvonne and {Rieke}, George and {Smith}, J.~D.~T. and {Whittet}, D.~C.~B.},
        title = "{Milky Way Mid-Infrared Spitzer Spectroscopic Extinction Curves: Continuum and Silicate Features}",
      journal = {\apj},
         year = 2021,
        month = jul,
       volume = {916},
       number = {1},
          eid = {33},
        pages = {33},
          doi = {10.3847/1538-4357/ac00b7},
archivePrefix = {arXiv},
       eprint = {2105.05087},
 primaryClass = {astro-ph.GA},
       adsurl = {https://ui.adsabs.harvard.edu/abs/2021ApJ...916...33G}
}

@ARTICLE{Gordon2023,
       author = {{Gordon}, Karl D. and {Clayton}, Geoffrey C. and {Decleir}, Marjorie and {Fitzpatrick}, E.~L. and {Massa}, Derck and {Misselt}, Karl A. and {Tollerud}, Erik J.},
        title = "{One Relation for All Wavelengths: The Far-ultraviolet to Mid-infrared Milky Way Spectroscopic R(V)-dependent Dust Extinction Relationship}",
      journal = {\apj},
         year = 2023,
        month = jun,
       volume = {950},
       number = {2},
          eid = {86},
        pages = {86},
          doi = {10.3847/1538-4357/accb59},
archivePrefix = {arXiv},
       eprint = {2304.01991},
 primaryClass = {astro-ph.GA},
       adsurl = {https://ui.adsabs.harvard.edu/abs/2023ApJ...950...86G}
}

@ARTICLE{Decleir2022,
       author = {{Decleir}, Marjorie and {Gordon}, Karl D. and {Andrews}, Jennifer E. and {Clayton}, Geoffrey C. and {Cushing}, Michael C. and {Misselt}, Karl A. and {Pendleton}, Yvonne and {Rayner}, John and {Vacca}, William D. and {Whittet}, D.~C.~B.},
        title = "{SpeX Near-infrared Spectroscopic Extinction Curves in the Milky Way}",
      journal = {\apj},
         year = 2022,
        month = may,
       volume = {930},
       number = {1},
          eid = {15},
        pages = {15},
          doi = {10.3847/1538-4357/ac5dbe},
archivePrefix = {arXiv},
       eprint = {2204.13716},
 primaryClass = {astro-ph.GA},
       adsurl = {https://ui.adsabs.harvard.edu/abs/2022ApJ...930...15D}
}

@ARTICLE{Gorski2005,
       author = {{G{\'o}rski}, K.~M. and {Hivon}, E. and {Banday}, A.~J. and {Wandelt}, B.~D. and {Hansen}, F.~K. and {Reinecke}, M. and {Bartelmann}, M.},
        title = "{HEALPix: A Framework for High-Resolution Discretization and Fast Analysis of Data Distributed on the Sphere}",
      journal = {\apj},
         year = 2005,
        month = apr,
       volume = {622},
       number = {2},
        pages = {759-771},
          doi = {10.1086/427976},
archivePrefix = {arXiv},
       eprint = {astro-ph/0409513},
 primaryClass = {astro-ph},
       adsurl = {https://ui.adsabs.harvard.edu/abs/2005ApJ...622..759G}
}

@ARTICLE{Kelsall1998,
       author = {{Kelsall}, T. and {Weiland}, J.~L. and {Franz}, B.~A. and {Reach}, W.~T. and {Arendt}, R.~G. and {Dwek}, E. and {Freudenreich}, H.~T. and {Hauser}, M.~G. and {Moseley}, S.~H. and {Odegard}, N.~P. and {Silverberg}, R.~F. and {Wright}, E.~L.},
        title = "{The COBE Diffuse Infrared Background Experiment Search for the Cosmic Infrared Background. II. Model of the Interplanetary Dust Cloud}",
      journal = {\apj},
         year = 1998,
        month = nov,
       volume = {508},
       number = {1},
        pages = {44-73},
          doi = {10.1086/306380},
archivePrefix = {arXiv},
       eprint = {astro-ph/9806250},
 primaryClass = {astro-ph},
       adsurl = {https://ui.adsabs.harvard.edu/abs/1998ApJ...508...44K}
}

@ARTICLE{Hensley2022,
       author = {{Hensley}, Brandon S. and {Murray}, Claire E. and {Dodici}, Mark},
        title = "{Polycyclic Aromatic Hydrocarbons, Anomalous Microwave Emission, and their Connection to the Cold Neutral Medium}",
      journal = {\apj},
         year = 2022,
        month = apr,
       volume = {929},
       number = {1},
          eid = {23},
        pages = {23},
          doi = {10.3847/1538-4357/ac5cbd},
archivePrefix = {arXiv},
       eprint = {2111.03067},
 primaryClass = {astro-ph.GA},
       adsurl = {https://ui.adsabs.harvard.edu/abs/2022ApJ...929...23H}
}

@ARTICLE{Schlegel1998,
       author = {{Schlegel}, David J. and {Finkbeiner}, Douglas P. and {Davis}, Marc},
        title = "{Maps of Dust Infrared Emission for Use in Estimation of Reddening and Cosmic Microwave Background Radiation Foregrounds}",
      journal = {\apj},
         year = 1998,
        month = jun,
       volume = {500},
       number = {2},
        pages = {525-553},
          doi = {10.1086/305772},
archivePrefix = {arXiv},
       eprint = {astro-ph/9710327},
 primaryClass = {astro-ph},
       adsurl = {https://ui.adsabs.harvard.edu/abs/1998ApJ...500..525S}
}

@ARTICLE{Egorov2023,
       author = {{Egorov}, Oleg V. and {Kreckel}, Kathryn and {Sandstrom}, Karin M. and {Leroy}, Adam K. and {Glover}, Simon C.~O. and {Groves}, Brent and {Kruijssen}, J.~M. Diederik and {Barnes}, Ashley. T. and {Belfiore}, Francesco and {Bigiel}, F. and {Blanc}, Guillermo A. and {Boquien}, M{\'e}d{\'e}ric and {Cao}, Yixian and {Chastenet}, J{\'e}r{\'e}my and {Chevance}, M{\'e}lanie and {Congiu}, Enrico and {Dale}, Daniel A. and {Emsellem}, Eric and {Grasha}, Kathryn and {Klessen}, Ralf S. and {Larson}, Kirsten L. and {Liu}, Daizhong and {Murphy}, Eric J. and {Pan}, Hsi-An and {Pessa}, Ismael and {Pety}, J{\'e}r{\^o}me and {Rosolowsky}, Erik and {Scheuermann}, Fabian and {Schinnerer}, Eva and {Sutter}, Jessica and {Thilker}, David A. and {Watkins}, Elizabeth J. and {Williams}, Thomas G.},
        title = "{PHANGS-JWST First Results: Destruction of the PAH Molecules in H II Regions Probed by JWST and MUSE}",
      journal = {\apjl},
         year = 2023,
        month = feb,
       volume = {944},
       number = {2},
          eid = {L16},
        pages = {L16},
          doi = {10.3847/2041-8213/acac92},
archivePrefix = {arXiv},
       eprint = {2212.09159},
 primaryClass = {astro-ph.GA},
       adsurl = {https://ui.adsabs.harvard.edu/abs/2023ApJ...944L..16E}
}

@ARTICLE{Egorov2025,
       author = {{Egorov}, Oleg V. and {Leroy}, Adam K. and {Sandstrom}, Karin and {Kreckel}, Kathryn and {Baron}, Dalya and {Belfiore}, Francesco and {Chown}, Ryan and {Sutter}, Jessica and {Boquien}, M{\'e}d{\'e}ric and {Canal i Saguer}, Mar and {Congiu}, Enrico and {Dale}, Daniel A. and {Egorova}, Evgeniya and {Huber}, Michael and {Li}, Jing and {Williams}, Thomas G. and {Chastenet}, J{\'e}r{\'e}my and {Chiang}, I.-Da and {Gerasimov}, Ivan and {Hassani}, Hamid and {Kim}, Hwihyun and {Koziol}, Hannah and {Lee}, Janice C. and {McClain}, Rebecca L. and {Delgado}, Jos{\'e} Eduardo M{\'e}ndez and {Pan}, Hsi-An and {Pathak}, Debosmita and {Rosolowsky}, Erik and {Sarbadhicary}, Sumit K. and {Schinnerer}, Eva and {Thilker}, David and {Ubeda}, Leonardo and {Weinbeck}, Tony},
        title = "{Polycyclic aromatic hydrocarbon destruction in star-forming regions across 42 nearby galaxies}",
      journal = {\aap},
         year = 2025,
        month = nov,
       volume = {703},
          eid = {A103},
        pages = {A103},
          doi = {10.1051/0004-6361/202556427},
archivePrefix = {arXiv},
       eprint = {2509.13845},
 primaryClass = {astro-ph.GA},
       adsurl = {https://ui.adsabs.harvard.edu/abs/2025A&A...703A.103E}
}

@ARTICLE{Lee2025,
       author = {{Lee}, Dennis and {Hensley}, Brandon S. and {Chang}, Tzu-Ching and {Dor{\'e}}, Olivier},
        title = "{The End of the Road for Far-infrared Reddening Maps? Evidence for Reddening Errors Driven by Changes in Polycyclic Aromatic Hydrocarbon Abundance}",
      journal = {\apj},
         year = 2025,
        month = nov,
       volume = {994},
       number = {1},
          eid = {61},
        pages = {61},
          doi = {10.3847/1538-4357/ae0c12},
archivePrefix = {arXiv},
       eprint = {2508.11746},
 primaryClass = {astro-ph.GA},
       adsurl = {https://ui.adsabs.harvard.edu/abs/2025ApJ...994...61L}
}

@ARTICLE{Yang2016,
       author = {{Yang}, X.~J. and {Li}, Aigen and {Glaser}, R. and {Zhong}, J.~X.},
        title = "{The C-H Stretching Features at 3.2--3.5 {\ensuremath{\mu}}m of Polycyclic Aromatic Hydrocarbons with Aliphatic Sidegroups}",
      journal = {\apj},
         year = 2016,
        month = jul,
       volume = {825},
       number = {1},
          eid = {22},
        pages = {22},
          doi = {10.3847/0004-637X/825/1/22},
archivePrefix = {arXiv},
       eprint = {1608.06704},
 primaryClass = {astro-ph.GA},
       adsurl = {https://ui.adsabs.harvard.edu/abs/2016ApJ...825...22Y}
}

@ARTICLE{Whitcomb2024,
       author = {{Whitcomb}, Cory M. and {Smith}, J.-D.~T. and {Sandstrom}, Karin and {Starkey}, Carl A. and {Donnelly}, Grant P. and {Draine}, Bruce T. and {Skillman}, Evan D. and {Dale}, Daniel A. and {Armus}, Lee and {Hensley}, Brandon S. and {Lai}, Thomas S.-Y. and {Kennicutt}, Robert C.},
        title = "{The Metallicity Dependence of PAH Emission in Galaxies. I. Insights from Deep Radial Spitzer Spectroscopy}",
      journal = {\apj},
         year = 2024,
        month = oct,
       volume = {974},
       number = {1},
          eid = {20},
        pages = {20},
          doi = {10.3847/1538-4357/ad66c8},
archivePrefix = {arXiv},
       eprint = {2405.09685},
 primaryClass = {astro-ph.GA},
       adsurl = {https://ui.adsabs.harvard.edu/abs/2024ApJ...974...20W}
}

@ARTICLE{Pare2026,
       author = {{Par{\'e}}, Dylan M. and {Chuss}, David T. and {Sponseller}, Danielle and {Hensley}, Brandon and {Kogut}, Alan},
        title = "{Analyzing Polycyclic Aromatic Hydrocarbons as a Tracer of Anomalous Microwave Emission near the Galactic Plane Using the COSMOGLOBE DIRBE Reduction}",
      journal = {\apj},
         year = 2026,
        month = feb,
       volume = {998},
       number = {2},
          eid = {337},
        pages = {337},
          doi = {10.3847/1538-4357/ae3f22},
archivePrefix = {arXiv},
       eprint = {2601.20545},
 primaryClass = {astro-ph.GA},
       adsurl = {https://ui.adsabs.harvard.edu/abs/2026ApJ...998..337P}
}

@ARTICLE{Richie2025,
       author = {{Richie}, Helena M. and {Hensley}, Brandon S.},
        title = "{PAH Emission Spectra and Band Ratios for Arbitrary Radiation Fields with the Single Photon Approximation}",
      journal = {arXiv e-prints},
         year = 2025,
        month = oct,
          eid = {arXiv:2510.16861},
        pages = {arXiv:2510.16861},
          doi = {10.48550/arXiv.2510.16861},
archivePrefix = {arXiv},
       eprint = {2510.16861},
 primaryClass = {astro-ph.GA},
       adsurl = {https://ui.adsabs.harvard.edu/abs/2025arXiv251016861R}
}

@ARTICLE{Planck2011,
author = {{\sorthelp{Planck Collaboration 2011T}}{Planck Collaboration XX}},
title = "{\textit{Planck} early results. XX. New light on anomalous microwave
 emission from spinning dust grains}",
journal = {\aap},
archivePrefix = "arXiv",
eprint = {1101.2031},
year = 2011,
volume = 536,
pages = {A20},
doi = {10.1051/0004-6361/201116470}
}

@ARTICLE{Planck2016,
author = {{\sorthelp{Planck Collaboration IntZW}}{Planck Collaboration Int.
 XLVIII}},
title = "{\textit{Planck} intermediate results. XLVIII. Disentangling Galactic
 dust emission and cosmic infrared background anisotropies}",
journal = {\aap},
archivePrefix = "arXiv",
eprint = {1605.09387},
year = 2016,
volume = 596,
pages = {A109},
doi = {10.1051/0004-6361/201629022}
}

@ARTICLE{Korngut2026,
       author = {{Korngut}, Phil M. and {Bock}, James J. and {Condon}, Samuel and {Dowell}, C. Darren and {Fazar}, Candice M. and {Hui}, Howard and {Moore}, Bradley D. and {Naylor}, Bret J. and {Nguyen}, Chi H. and {Padin}, Stephen and {Wincentsen}, James and {Aboobaker}, Asad M. and {Akeson}, Rachel and {Alred}, John M. and {Alibay}, Farah and {Ashby}, Matthew L.~N. and {Bach}, Yoonsoo P. and {Bichel}, Joseph and {Bolton}, Douglas and {Braun}, David F. and {Brown}, Thomas and {Bryan}, Sean A. and {Burnham}, Jill and {Burk}, Thomas A. and {Burke}, Nicholas and {Catching}, Ben and {Chang}, Tzu-Ching and {Chen}, Shuang-Shuang and {Cheng}, Yun-Ting and {Chiang}, Yi-Kuan and {Chong}, Yong and {Cooray}, Asantha and {Cook}, Walter R. and {Cormarkovic}, Velibor and {Crill}, Brendan P. and {Cukierman}, Ari J. and {Davis}, Andrew and {Darga}, Dan and {Disarro}, Thomas and {Dore}, Olivier and {Fabinsky}, Beth E. and {Faisst}, Andreas L. and {Fanson}, James L. and {Farrington}, Allen H. and {Fatahi}, Tamim and {Feder}, Richard M. and {Frater}, Eric H. and {Goldina}, Tatiana and {Gorjian}, Varoujan and {Hart}, William G. and {Hendricks}, Warren and {Hora}, Joseph L. and {Hristov}, Viktor and {Huai}, Zhaoyu and {Hulse}, Charles A. and {Jo}, Young-Soo and {Jeong}, Woong-Seob and {Kamei}, Makenzie L. and {Kang}, Jae Hwan and {Kecman}, Branislav and {Marchant}, Will and {Mariani}, Giacomo and {Masters}, Daniel C. and {Melnick}, Gary J. and {Miyasaka}, Hiromasa and {Murgia}, Giulia and {Nelson}, Christina and {Nguyen}, Hien T. and {Owen}, Christopher and {Paladini}, Roberta and {Park}, Sung-Joon and {Patil}, Harshad and {Penanen}, Konstantin and {Piazzo}, Chris and {Pyo}, Jeonghyun and {Quon}, Amelia and {Ramanathan}, Keshav and {Rustamkulov}, Zafar and {Reiley}, Daniel J. and {Rice}, Eric B. and {Ridenhour}, Flora and {Roberts}, Amber and {Rocca}, Jennifer M. and {Signorini}, Alessandro and {Susca}, Sara and {Tolls}, Volker and {Velicheti}, Phani and {Wang}, Pao-Yu and {Werner}, Michael W. and {White}, Casey and {Williamson}, Ross and {Yang}, Yujin and {Zemcov}, Michael},
        title = "{The SPHEREx Instrument: Calibration, testing and performance measurements of the NIR 2 spectroscopic surveyor from the laboratory to in-orbit commissioning}",
      journal = {arXiv e-prints},
         year = 2026,
        month = mar,
          eid = {arXiv:2603.29835},
        pages = {arXiv:2603.29835},
          doi = {10.48550/arXiv.2603.29835},
archivePrefix = {arXiv},
       eprint = {2603.29835},
 primaryClass = {astro-ph.IM},
       adsurl = {https://ui.adsabs.harvard.edu/abs/2026arXiv260329835K}
}

@article{Hui_2026,
    year = {in prep.},
    author = {Howard Hui and others}
}

@article{Yang2026,
    year = {in prep.},
    author = {Yujin Yang and others}
}

@ARTICLE{Cukierman2026,
       author = {{Cukierman}, Ari J. and {Chen}, Shuang-Shuang and {Kang}, Jae Hwan and {Minasyan}, Mary H. and {Murgia}, Giulia and {Bock}, James J. and {Chang}, Tzu-Ching and {Chiang}, Yi-Kuan and {Crill}, Brendan P. and {Dor{\'e}}, Olivier and {Dowell}, C. Darren and {Faisst}, Andreas L. and {Hora}, Joseph L. and {Hui}, Howard and {Kang}, Miju and {Korngut}, Phil M. and {Lee}, Ho-Gyu and {Lee}, Bomee and {Melnick}, Gary J. and {Mirocha}, Jordan and {Nguyen}, Chi H. and {Rustamkulov}, Zafar and {Tolls}, Volker and {Werner}, Michael W. and {Yang}, Yujin and {Zemcov}, Michael},
        title = "{Spectral Map Making with SPHEREx}",
      journal = {arXiv e-prints},
         year = 2026,
        month = mar,
          eid = {arXiv:2603.25790},
        pages = {arXiv:2603.25790},
          doi = {10.48550/arXiv.2603.25790},
archivePrefix = {arXiv},
       eprint = {2603.25790},
 primaryClass = {astro-ph.IM},
       adsurl = {https://ui.adsabs.harvard.edu/abs/2026arXiv260325790C}
}

@misc{SPHERExQR2,
doi = {10.26131/IRSA652},
url = {https://catcopy.ipac.caltech.edu/dois/doi.php?id=10.26131/IRSA652},
author = {{SPHEREx Team}},
title = {SPHEREx Quick Release Spectral Images - QR2},
publisher = {IPAC},
year = {2025}
}
\bibliographystyle{aasjournalv7}

\end{document}